\documentclass[final,5p,times,twocolumn]{elsarticle}
\UseRawInputEncoding
\usepackage{lineno}
\modulolinenumbers[5]
\journal{the Journal of Information and Software Technology}
\usepackage{subfig}
\usepackage{array}
\usepackage{lscape}
\usepackage{multirow}
\usepackage{graphicx}
\usepackage{rotating}
\usepackage{makecell}
\usepackage{booktabs, tabularx}
\usepackage{geometry}
\usepackage{graphicx}
\usepackage{ltablex}
\usepackage{amsmath}
\usepackage[tableposition=t]{caption}
\newcolumntype{L}{>{\raggedright\arraybackslash}X}
\usepackage{comment}
\usepackage{afterpage}
\usepackage{enumitem}%
\usepackage{multicol}
\setlist{nosep, noitemsep}
\usepackage{enumitem}
\usepackage{comment}
\usepackage{booktabs} 
\usepackage{pdfpages}
\usepackage{soul}
\usepackage{pdflscape}
\usepackage{subfig}
\usepackage{fontawesome}
\usepackage{tabularx}
\usepackage{makecell}
\usepackage{tcolorbox}
\usepackage{dirtytalk}
\usepackage{float}
\usepackage{multirow}
\usepackage{supertabular}
\usepackage{svg}
\usepackage[framemethod=tikz]{mdframed}
\usepackage[hyphens]{url}
\usepackage{hyperref}
\usepackage[hyphenbreaks]{breakurl}

\usepackage{xcolor}

\usepackage{etoolbox}
\makeatletter
\patchcmd{\ps@pprintTitle}
  {Preprint submitted}
  {Submitted}
  {}{}
\makeatother

\begin{document}

\begin{frontmatter}

\title{Challenges and solutions when adopting DevSecOps: A systematic review}

\author[mymainaddress,mysecondaryaddress]{Roshan N. Rajapakse\corref{correspondingauthor}} 
\ead{roshan.rajapakse@adelaide.edu.au}
\author[mymainaddress]{Mansooreh Zahedi} \ead{mansooreh.zahedi@adelaide.edu.au}
\author[mymainaddress,mysecondaryaddress]{M. Ali Babar} \ead{ali.babar@adelaide.edu.au}
\author[mytertiaryaddress]{Haifeng Shen} \ead{Haifeng.Shen@acu.edu.au}


\cortext[correspondingauthor]{Corresponding author}

\address[mymainaddress]{Centre for Research on Engineering Software Technologies, School of Computer Science, University of Adelaide, Adelaide}
\address[mysecondaryaddress]{Cyber Security Cooperative Research Centre, Australia}
\address[mytertiaryaddress]{The HilstLab, Peter Faber Business School, Australian Catholic University, Sydney, Australia}

\begin{abstract}

\textbf{Context:} DevOps (Development and Operations) has become one of the fastest-growing software development paradigms in the industry. However, this trend has presented the challenge of ensuring secure software delivery while maintaining the agility of DevOps. The efforts to integrate security in DevOps have resulted in the DevSecOps paradigm, which is gaining significant interest from both industry and academia. However, the adoption of DevSecOps in practice is proving to be a challenge.

\textbf{Objective:} This study aims to systemize the knowledge about the challenges faced by practitioners when adopting DevSecOps and the proposed solutions reported in the literature. We also aim to identify the areas that need further research in the future.

\textbf{Method:} We conducted a Systematic Literature Review of 54 peer-reviewed studies. The thematic analysis method was applied to analyze the extracted data.

\textbf{Results:} We identified 21 challenges related to adopting DevSecOps, 31 specific solutions, and the mapping between these findings. We also determined key gap areas in this domain by holistically evaluating the available solutions against the challenges. The results of the study were classified into four themes: People, Practices, Tools, and Infrastructure. Our findings demonstrate that tool-related challenges and solutions were the most frequently reported, driven by the need for automation in this paradigm. Shift-left security and continuous security assessment were two key practices recommended for DevSecOps. People-related factors were considered critical for successful DevSecOps adoption but less studied.

\textbf{Conclusions:} We highlight the need for developer-centered application security testing tools that target the continuous practices in DevSecOps. More research is needed on how the traditionally manual security practices can be automated to suit rapid software deployment cycles. Finally, achieving a suitable balance between the speed of delivery and security is a significant issue practitioners face in the DevSecOps paradigm.

\end{abstract}


\end{frontmatter}


\section{Introduction}

DevOps (\textbf{Dev}elopment and \textbf{Op}eration\textbf{s}) has led to a paradigm shift aimed at removing the traditional boundaries (or \say{silos}) of the software development and software operations teams \cite{leite2019survey}. This shift resulted in reducing the time between committing a modification in a system and that change being placed in a production environment \cite{bass2015devops}. DevOps is currently a widely adopted software development paradigm in the industry \cite{mann2018state}. This interest in adoption is due to the gains in business value reported by industry practitioners and academic researchers \cite{SignalSciences2020}. The most commonly reported benefit is the ability to deploy releases faster and more frequently \cite{riungu2016devops}. However, the practices of rapid delivery have presented new challenges to organizations. One such challenge is ensuring the security of software outputs to stakeholders while maintaining the agility of DevOps \cite{myrbakken2017devsecops}. 

Traditionally, security is treated as a non-functional requirement \cite{flechais2005designing}, which is handled at a later stage of the software development life-cycle \cite{Shiftleft2021}, \cite{Sharma2020}. Accordingly, a set of standard application security tests or activities are conducted on a software release. These activities either need substantial manual effort (e.g., security code review \cite{howard2006process}) or are time consuming tasks (e.g., Dynamic Application Security Testing (DAST) \cite{peterson2020}). Therefore, applying the same security tests in the context of DevOps would hinder the speed of deployments. At the same time, with the rising number of attacks, the security of software is critical in today's context, particularly in a cloud environment. There are many examples of how security vulnerabilities of software have been exploited to cause substantial damages to organizations \cite{Ng2018}, \cite{Bushwick2020}. As a result, a key focus in this domain is how an organization can produce outputs at the speed required by DevOps while ensuring security \cite{mann2019state}.

\begin{figure*}[t!]
    
    \centering
    \includegraphics [width=.8\textwidth]{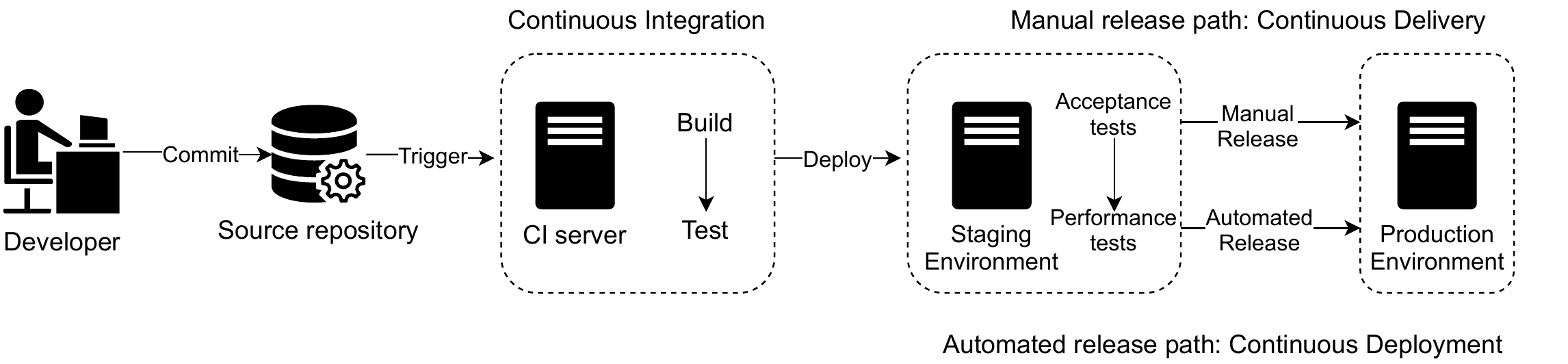}
    \caption{The relationships between the continuous practices discussed in this study \cite{Prince2016}, \cite{shahin2017continuous}, \cite{shahin2019empirical}}
    \label{fig:CI-CD-CDE}
\end{figure*}

This requirement of integrating security in DevOps has led to the coining of the term, DevSecOps (Development, \textbf{Sec}urity, and Operations). At the core of DevSecOps is the principle of keeping security as a priority and adding security controls and practices into the DevOps cycle \cite{myrbakken2017devsecops}. As the need for rapid deployment of safe and secure software outputs increases, the interest in DevSecOps continues to grow in the industry and academia. This is evident from the growing body of formal literature in this area. 

\subsection{Aim and contribution}

The transition from traditional software engineering methodologies (e.g., waterfall model) to DevSecOps is widely reported to be a challenging task \cite{SignalSciences2020}, \cite{checkmarx_2020}, \cite{mann2019state}. As the interest in DevSecOps continues to rise in the industry, it is valuable for practitioners to be aware of such adoption challenges and the solutions available to address them. 

Firstly, this article is intended for practitioners who are planning or in the process of adopting DevSecOps to be aware of the frequently reported problems in this domain. Identifying the adoption challenges at a very early stage of a project would be beneficial in addressing them early. Secondly, we aim to provide practitioners with a critical review of the proposed solutions related to DevSecOps adoption, reported in the peer-reviewed studies. Thirdly, this study can be a starting point for further research in the research community, as we identify the gap areas in DevSecOps based on the current literature.

Accordingly, our main aim in this study is to systematically select, thematically analyze and present the challenges, solutions, and gaps for further research on DevSecOps. To achieve this aim, we have conducted an SLR to evaluate a selected set of peer-reviewed literature. Based on the results, our paper makes the following three specific contributions:
\newline
\begin{itemize}

\item We present a thematic classification of the main security-related challenges an organization could face in adopting DevSecOps.

\item We describe the current solutions proposed in the literature, which address these challenges in terms of guidelines, best practice, methodologies or frameworks, and tools. We then thematically map the challenges to the proposed solutions.

\item We identify the potential gaps for future research or the areas for technological development (e.g., tools) or framework support by combining the findings of the above two contributions.

\end{itemize}   

\section{Background and related work}

In this section, we define the terms and concepts used in the study. Then, we present a comparison of our study with the existing related reviews.

\subsection{Continuous software engineering and its practices}
Continuous Software Engineering (CSE) aims to establish a continuous movement in software engineering activities, rather than a set of discrete activities performed by different teams or departments \cite{fitzgerald2017continuous}. To enable this continuous movement, CSE bundles a set of continuous practices such as Continuous Integration. Fitzgerald and Stol \cite{fitzgerald2017continuous} present several such continuous practices categorized under business strategy \& planning, development, and operations. In the industry, the development-related continuous practices have been used more and are well-established \cite{bosch2014continuous}. Accordingly, much of the research on continuous practices has been conducted on the development-related practices \cite{shahin2017continuous}, \cite{schermann2016empirical}, \cite{shahin2019empirical}, \cite{zahedi2020mining}. Based on these studies, Continuous Integration, Continuous Delivery, and Continuous Deployment can be seen as the most popular continuous practices in the domain \cite{stahl2017continuous}. Therefore, we have selected these three practices for review in our study. In the next subsections, we present the definitions and relationships between these practices.

\subsubsection{Continuous Integration (CI)}
CI is the development practice where developers frequently integrate their work (e.g., code changes) to the main branch, usually on a daily basis \cite{stahl2017continuous}. These changes are validated by automated builds and tests \cite{leppanen2015highways}. By doing so, developers are able to detect and address integration failures early, and as quickly as possible \cite{shahin2017continuous}, \cite{fitzgerald2017continuous}.

\subsubsection{Continuous Delivery (CDE)}
CDE aims to keep the software at a reliably deployable or \textit{production-ready} state at any time \cite{chen2015continuous}, \cite{shahin2017continuous}. To achieve this goal, software needs to pass the relevant tests and quality checks in a staging environment (e.g., acceptance tests). However, the deployment to the production environment is done manually, where a team member with the relevant authority decides when and which production-ready outputs should be released to the customer (i.e., pull-based approach) \cite{shahin2019empirical}.

\subsubsection{Continuous Deployment (CD)}
CD extends CDE by automatically and continuously deploying software outputs to a production environment if all the required quality gates are passed [P18], \cite{shahin2017continuous}. CD is a push-based approach, where software changes are automatically deployed to production through the deployment pipeline without human intervention \cite{shahin2019empirical}. Figure \ref{fig:CI-CD-CDE} depicts the relationship between these continuous practices, which we have considered in our study.

\subsection{Development and Operations (DevOps)}
DevOps is a paradigm that aimed to reduce the disconnect between development and operations teams by promoting collaboration, communication, and integration between them \cite{bass2015devops}. Despite being a trend in the industry, DevOps lacks a widely accepted definition \cite{leite2019survey}. Defining DevOps is difficult due to the considerable overlap with continuous practices \cite{stahl2017continuous}. Due to this reason, Stahl et al. \cite{stahl2017continuous} present definitions to differentiate DevOps and continuous practices from one another. We use their interpretation for DevOps in our study, as they have attempted to reflect the mainstream interpretation of these terms. 

Stahl et al. \cite{stahl2017continuous} view DevOps as a combination of values, principles, methods, practices, and tools. Here, the practices include continuous activities such as CI, CD, and CDE. These are considered key practices which enable achieving the above-noted goal of DevOps \cite{bass2015devops}. The above five elements in combination, enable continuous and rapid delivery of software without compromising on the quality. 
\subsection{Development, Security and Operations (DevSecOps)}
The paradigm of DevSecOps refers to the integration of security principles and practices in DevOps through increased communication, collaboration, and integration between the development and operations teams with the security team [P42]. Studies and industry personnel have used variations of this term, such as SecDevOps \cite{mohan2016secdevops} and DevOpsSec \cite{bird2016devopssec}. However, the main intended goal in each of these terms is to keep security as a key focus throughout the DevOps cycle. 

\subsection{Other reviews in DevSecOps}

\begin{table*}[hbt!]
\centering
\caption{Comparison of other reviews in DevSecOps : *PC: Peer-reviewed literature count, *GC: Grey literature count, *OC: Overlapped peer-reviewed paper count with other reviews, *A: Did the study include security of continuous practices?, *B: Did the study include adoption challenges?, *C: Did the study include proposed solutions?, *D: Did the study include analysis of gap areas? [N/A: Not applicable] \newline \newline  \faCheck-  The authors have considered the security of continuous practices in the study.  \newline \faCheckSquare-  Adoption challenges, solutions or gap areas are addressed directly through research questions in the study. \newline \faCheckSquareO- Adoption challenges, solutions or gap areas are addressed partially through other research questions (i.e., by addressing a different research question, partial information related to challenges or solutions is provided.)}

\footnotesize
\begin{tabular}{ |p{3cm}|p{6cm}|p{.5cm}|p{.5cm}|p{.5cm}|p{.5cm}|p{.2cm}| p{.2cm}| p{.2cm}| p{.2cm}|}

\hline
\textbf{Authors}& \textbf{Focus} &\textbf{Year} &\textbf{PC} & \textbf{GC} & \textbf{OC} & \textbf{A} & \textbf{B} & \textbf{C} & \textbf{D}\\ 
\hline
Mohan \& ben Othmane \cite{mohan2016secdevops}& Definition, security best practice, compliance, process automation, tools, software configuration, team collaboration, availability of activity data and information secrecy related to DevSecOps & 2016 & 5 & 3 & 1 &  & \faCheckSquareO & \faCheckSquareO & \\ 
\hline
Myrbakken \& Colomo-Palacios \cite{myrbakken2017devsecops}& Definition, characteristics, adoption challenges, benefits and evolution of DevSecOps & 2017 & 2 & 50 & 0 &  & \faCheckSquare &  &  \\
\hline
L. Prates et al. \cite{prates2019devsecops}& DevSecOps Metrics & 2019 & 2 & 11 & 0 & &  & \faCheckSquareO & \\
\hline

S\'anchez-Gord\'on \& Colomo-Palacios \cite{sanchez2020security}& Characterizing DevSecOps culture & 2020 & 11 & N/A & 2 & & & \faCheckSquareO & \\
\hline
Mao et al. \cite{mao2020preliminary}& Impact of DevOps on software security, practitioners’ perceptions and practices associated with DevSecOps & 2020 & N/A & 141 & N/A & & & \faCheckSquareO & \\

\hline
\hline
Rafi et al. \cite{rafi2020prioritization}& Prioritization based taxonomy of DevOps security challenges & 2020 & 40 & N/A & 6 &  & \faCheckSquare &  & \\
\hline
\hline
\textbf{This study}& DevSecOps adoption challenges, solutions and gap areas & 2021 & 54 & N/A & N/A & \faCheck & \faCheckSquare & \faCheckSquare & \faCheckSquare \\
\hline
\end{tabular}
\label{table:other_reviews}
\end{table*}
\normalsize

We are aware that there are existing literature review and mapping studies related to security in DevOps or DevSecOps (Table \ref{table:other_reviews}).  In this section, we highlight the contribution of our SLR compared with these existing reviews.

The early review studies in DevSecOps have only included a limited number of resources, as this paradigm was relatively new in that period. For example, the systematic mapping study by Mohan and ben Othmane \cite{mohan2016secdevops}, which was conducted in 2016, contained only five peer-reviewed articles and three DevSecOps presentations from conferences. The authors aimed to determine whether DevSecOps or SecDevOps was just a buzzword during that period. Accordingly, Mohan and ben Othmane \cite{mohan2016secdevops} identified certain aspects related to DevSecOps noted in Table \ref{table:other_reviews}. Additionally, they discussed a limited number of challenges and solutions related to DevSecOps in some of the noted aspects. We provide more detailed coverage and a substantially larger number of DevSecOps adoption challenges and solutions in our study. 

Another early Multivocal Literature Review (MLR) by Myrbakken and Colomo-Palacios \cite{myrbakken2017devsecops} mostly relied on grey literature (e.g., white papers, blogs, and articles). They provided a definition, characteristics, benefits, adoption challenges, and evolution of DevSecOps. The authors also noted that the challenges were related to traditional or manual security methods, organizational problems, and lack of appropriate tools. By contrast, we provide more in-depth coverage in addressing the adoption challenges from the recent peer-reviewed literature.  

L. Prates et al. \cite{prates2019devsecops} aimed to uncover metrics that can be used to measure the effectiveness of the DevSecOps methodology using 13 sources. While our study's focus is not related to identifying specific metrics, we mention it as one aspect of measuring security in DevSecOps.

S\'anchez-Gord\'on and Colomo-Palacios \cite{sanchez2020security} conducted an SLR on DevSecOps, which included only 11 papers. In this study, the focus of the authors was on characterizing the culture in DevSecOps. Therefore, the main focus of this article is different from that of our study. Further, a Grey Literature Review (GLR) on DevSecOps was carried out by Mao et al. \cite{mao2020preliminary}. Our research differs from this study based on the research questions addressed. Their research questions investigated the impact of DevOps on software security, the aspects that practitioners use to understand DevSecOps and the key practices of the paradigm. Some of the practices captured by this study are reported as solutions in our study too. We map these practices to the challenges identified in our thematic analysis.

We have also identified studies that have used a systematic review of the literature as part of the study. The study conducted by Rafi et al. \cite{rafi2020prioritization} contained an SLR to identify the security challenges in DevOps as part of the research. The authors then evaluated the identified challenges using DevOps experts. However, only six papers overlap between this article and our study. We give reasons for this observation below.

Studies state that continuous practices such as CD are closely related with DevOps as a concept \cite{rahman2015synthesizing}. Based on the SLR by Stahl et al. \cite{stahl2017continuous}, many studies state that DevOps is enabled by continuous practices \cite{wettinger2015enabling}, \cite{gotimer2016devops}, \cite{olszewska2015devops}, \cite{wettinger2015dyn}. For example, CI, CD, and CDE are key practices that enable the rapid and continuous deployment cycles of the DevOps paradigm \cite{bass2015devops}, \cite{shahin2017beyond}. Another component of DevOps (as per Stahl et al. \cite{stahl2017continuous}) which is quite similar to continuous practices, is tools. Tools are a critical component in both continuous practices and the DevOps paradigm, as they enable automation.  We note that there are many overlaps between what practitioners consider as DevOps tools and tools used in continuous practices (e.g., CI tools) \cite{zahedi2020mining}. Based on the above reasons, we argue that to cover security of DevOps, security of these continuous practices needs to be considered.  

To verify this argument, we conducted pilot searches in digital libraries using the search terms that captured the relevant studies (e.g., security AND \say{continuous integration}). By analyzing the results, we found a number of studies that can contribute to our research questions. However, none of the previous SLRs or MLRs (Table \ref{table:other_reviews}) with similar research questions have considered the security of these continuous practices, which we have captured in our study. Only the study by Rahman and Williams [P42], which is not a core literature review study, has captured this aspect. When searching for internet artifacts in Google search to answer their research questions and then to prepare a survey on security in DevOps, they used \say{Security in Continuous Delivery} and \say{Security in Continuous Deployment} as search terms. 

Lastly, Stahl et al. \cite{stahl2017continuous} state that the terms DevOps and continuous practices are widely used interchangeably. Therefore, our decision to cover security of continuous practices reduces the possibility of missing out on the relevant studies. Ultimately, this is why we managed to capture more peer-reviewed studies relevant to security in DevOps than the previous SLRs or MLRs. In summary, our review differs from the existing studies in the following ways.
\newline
\begin{itemize}
    \item To our knowledge, the combination of challenges related to adopting DevSecOps and proposed solutions have not been systematically reviewed using a substantial body of literature. By identifying the challenges, solutions, and the mapping between them, we were able to identify key gap areas in this domain.
    \item We have considered security of the key continuous practices which enable DevOps as part of our study. This resulted in capturing a large set of relevant studies, which were not included in the previous studies.\newline
\end{itemize}

The rest of this paper is organized as follows. In Section 3, we present the research methodology used in this study. This is followed by the results and discussion in Section 4. In Section 5, we present the threats to validity and finally, Section 6 concludes our study.

\begin{table*}[t!]
\centering
\caption{Research questions addressed in this study.}
\small
\resizebox{\textwidth}{!}{%
\begin{tabular}{p{0.4\textwidth}p{0.6\textwidth}}
\\ \toprule
\textbf{Research questions}&\textbf{Motivation}\\
\midrule
\textbf{RQ1:} What are the specific challenges related to adopting DevSecOps reported in the previous research? & For organizations that are planning to adopt DevSecOps, early identification of the adoption challenges is important. However, a single peer-reviewed source that systematically analyzes and synthesizes these adoption problems using a significant body of up-to-date academic literature is unavailable. This research question aims to address this specific gap. \\
\textbf{RQ2:} What are the solutions proposed in the previous research to address the DevSecOps adoption challenges? & An aspect of DevSecOps which is of high interest for the practitioners is solutions in terms of specific guidelines, best practice, and tools, frameworks or technologies. This area has not been addressed rigorously by the previous reviews or survey studies in DevSecOps. \\
\textbf{RQ3:} What are the opportunities for future research or gap areas for technological development (e.g., tool support) or framework support in this domain?& As an area with a growing interest in the research community, it is important to identify the research gaps to plan future studies. By evaluating the findings of RQ1 and 2, we aim to identify such gap areas for the research community. Further, we also aim to highlight the areas in DevSecOps where tool or framework support is lacking. This would be useful information for the industry (e.g., tool vendors). \\

\bottomrule
\end{tabular}}
\label{table:RQs}
\normalsize
\end{table*}%

\section{Methodology}
SLRs are considered as one of the most popular methods in the field of Evidence-Based Software Engineering (EBSE) \cite{kitchenham2004evidence}. Studies note that SLRs can enable practitioners to make informed decisions related to technology selection and adoption \cite{dyba2005evidence}. Therefore, we decided to conduct an SLR to answer our research questions related to DevSecOps adoption. 

To carry out this study, we followed the SLR guidelines prepared by Kitchenham and Charters \cite{Kitchenham07guidelinesfor}.  Also, similar to the SLR on CD by Laukkanen et al. \cite{laukkanen2017problems}, we included multiple studies of the same project if new contributions were available. This was done in order to capture all the relevant information for our research questions. According to this guide \cite{Kitchenham07guidelinesfor}, the study design is presented in the below subsections: (i) Research questions, (ii) Search strategy, (iii) Inclusion and exclusion criteria, (iv) Data extraction, (v) Data synthesis. 

\subsection{Research questions}

We note the research questions and the motivation behind each question in Table \ref{table:RQs}. All the authors discussed and agreed on the research questions and the process of conducting the search. However, the first author conducted the search and filtered the studies under the close supervision of the other authors who are experienced researchers. These steps are detailed in the following sections.

\subsection{Search strategy}
We used the guide given by Kitchenham and Charters \cite{Kitchenham07guidelinesfor} to iteratively develop the search string of this study. Our search string consists of two types of search terms. Firstly, we selected the more popular term DevSecOps and other similar terms that are used interchangeably (e.g., SecDevOps). To find such terms, we consulted industry reports and previous reviews.

Secondly, as noted in Section 2.1, we aimed to capture studies that addressed the security issues of the widely researched continuous practices in DevOps. Therefore, as the second part of the search string, we selected search terms that would capture the security aspects of these continuous practices (e.g., secur* AND \say{continuous integration}). 

The final search string is presented below.

\begin{tcolorbox}[left=3pt, top=2pt, right=3pt, bottom=2pt]

devsecops OR secdevops OR devopssec OR secops OR \say{rugged devops} OR ruggedops OR (secur* AND devops) OR (secur* AND \say{continuous software engineering}) OR (secur* AND \say{continuous delivery}) OR (secur* AND \say{continuous deployment}) OR (secur* AND \say{continuous integration})

\end{tcolorbox}

We selected one index engine (i.e., Scopus) as studies have shown that there is a significant overlap among index engines \cite{chen2010towards}. Then, we selected two publisher sites (i.e., IEEE Xplore, ACM Digital Library) for our study.  Next, we conducted pilot searches on the three selected sources to check whether several key papers were included in the results. Our selections are further supported by Chen et al. \cite{chen2010towards} recommendations (e.g., IEEE Xplore and ACM Digital Library have considerable exclusive contributions, and both of these sources have significant overlap with Google Scholar). 

We ran the search in June 2020 on the selected sources. The search terms were matched only with the title, abstract, and keywords of papers. As a result, we retrieved a total of 460 papers. We then removed the duplicates and ended up with 283 papers.

\subsection{Inclusion-exclusion criteria}
We used the following inclusion and exclusion criteria to filter the papers resulting from the above step.

\textit{Inclusion criteria:}
\begin{itemize}
\item The article specifically addresses (e.g., study objective or research questions) some aspect of security in one or more of the stages of DevOps or the selected continuous practices.
\item The full text is available.
\item The article is written in English.
\end{itemize}

\textit{Exclusion criteria:}
\begin{itemize}
\item Publications that are not peer-reviewed (e.g., keynote abstracts, call for papers, and presentations)
\item The main element or contribution of the article is a literature review or survey (e.g., secondary study).
\item The article is a short paper (i.e., five pages or below) 
\end{itemize}

At first, we used the title and abstract for this purpose. However, in certain papers, we needed to consider the full text to make the decision. In these papers, even though our search string terms were present in the title, abstract, or keywords, it was unclear how the paper's content was related to the focus of our SLR. For example, in several papers, the term DevOps was noted in the title or abstract. However, upon inspecting the full paper, we did not find the content to be adequately related to DevOps (or continuous practices). To determine whether a paper was adequately related to the focus of our review (i.e., security in DevOps or continuous practices), we assessed whether the aim or objective of the study or the research questions explicitly addressed these areas. By following this step, we were able to remove studies where the relevant keywords (e.g., DevOps) were only included in the paper to provide background or attract readers (as these are areas with high interest) but not the main focus of the study.

The first author shared an Excel sheet that detailed the selection decisions for each included or excluded paper among the other authors, who conducted a detailed review. In the weekly meetings, the first author used this sheet to discuss the task in detail, and the other authors provided feedback which led to paper inclusions and exclusions. We carried out several rounds of discussions until a consensus was reached about the selected list of papers. At the end of this process, we selected 62 papers based on the inclusion and exclusion criteria.  

    \begin{figure*}[t!]
    
    \centering
    \includegraphics[origin=c, width=1\textwidth]{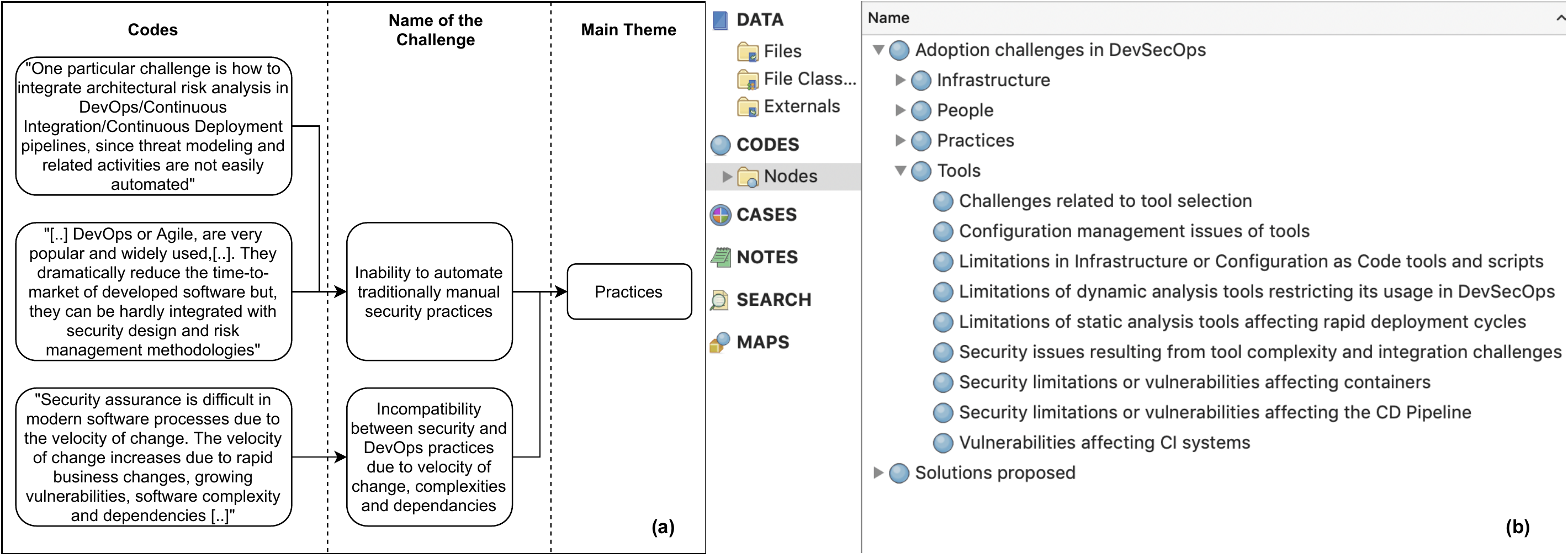}
    \caption{(a) Example of the coding process  (b) Multi-layered coding in NVivo}
    \label{fig:nvivo}
    
    \end{figure*}

\subsection{Quality assessment criteria}
We used the following quality assessment criteria (adopted from \cite{dybaa2008empirical}) to assess the papers resulting from the above stage. As our study contained a high amount of industry-authored papers, we limited our quality assessment for the below criteria.
\begin{itemize}
    \item \textit{Adequate description of the context:} The context or organizational settings of the paper (e.g., particularly in industry studies) are clearly noted.
    \item \textit{The research design or method suits the aims of the study:} The methodology followed is appropriate to address the research goals.
    \item \textit{Adequate description of the result or solution (i.e., findings):} Clear statement of the findings or contributions of the paper (e.g., solution proposal, tool) is presented.
\end{itemize}

The grading of these criteria was done using a binary scale (\textit{Yes/No}) in the previously noted Excel sheet by the first author and reviewed by the other authors. We removed a paper if it received more than one \textit{No} grade for the three criteria. After the quality assessment was done, we selected a total of 48 articles for this study. 

\subsection{Snowballing activity}
The first author conducted forward and backward snowballing on the 48 selected papers, which the other authors supervised. The instructions for this activity were obtained from Wohlin et al. \cite{wohlin2014guidelines}. We found that most of the papers with in-scope content were already captured in our search stage. As a result, only four papers were added using this activity.

We then performed snowballing on the primary study lists of the secondary studies noted in Table {\ref{table:other_reviews}}. This was done to identify any potentially missed primary studies. By screening a total of 49 papers from the primary study lists of these studies, we identified two papers that satisfied our inclusion/exclusion and quality assessment criteria. Accordingly, the final in-scope paper count for our study is 54 papers. The final set of articles is listed in the Appendix.

\subsection{Data extraction}

We adopted and modified the data extraction sheet from Garousi et al. \cite{garousi2019guidelines} to extract data of the papers into Microsoft Excel. Kitchenham and Charters \cite{Kitchenham07guidelinesfor} recommended that data extraction should be performed independently by two or more authors. Therefore, the first and fourth author (a senior researcher) performed the data extraction in our study. The first author extracted data from 39 papers, while the fourth author extracted from the remaining 15 papers. We used an excel function to select these 15 papers randomly. After the data extraction was completed, both authors further examined the extraction sheets to ensure consistency. These sheets were then shared with the remaining authors for review.

\subsection{Data synthesis and mapping}
The first author then carried out the data synthesis. The other authors regularly reviewed the extracted data and synthesis results for improvements and to resolve issues. 

As we aimed to classify the reported challenges and solutions in the domain, we decided to use the thematic analysis approach of qualitative research for our study. For this purpose, we followed the thematic analysis process reported by Braun and Clarke \cite{braun2006using} to synthesize the extracted data. 
The extracted data were imported into NVivo, a qualitative data analysis tool \cite{braun2006using}. We then performed \textit{open coding} on the extracted data using this tool. In open coding, the data is broken down into smaller components and labeled using a \textit{code} \cite{sbaraini2011grounded} (a code is a word or phrase that acts as a label for a selection of meaningful text \cite{codes_description}).
We conducted this activity in an iterative manner, where the codes assigned in the first attempt were modified in the later rounds (e.g., coded content and the name of the code). Finally, we analyzed the relationship between these codes and formed the overall themes of the study. Figure \ref{fig:nvivo} (a) presents an example of how the themes were formed using the coded content. For this process, we employed a multi-layered coding structure in NVivo as depicted in Figure \ref{fig:nvivo} (b).

   \begin{figure*}[t!]
    
    \centering
    \includegraphics[origin=c, width=1\textwidth]{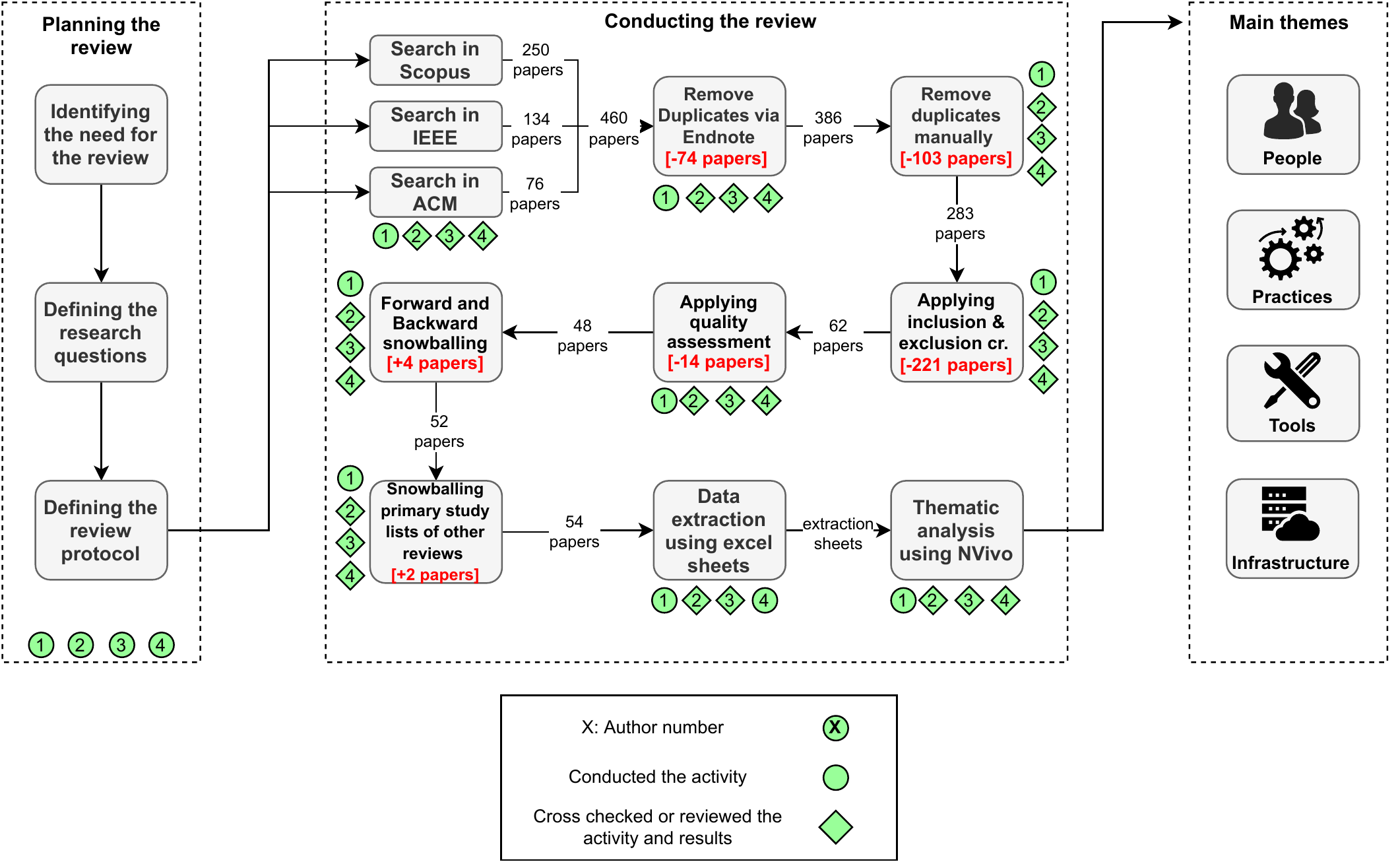}
    \caption{The overview of our research process}
    \label{fig:method}
    
    \end{figure*}

Further, we considered the advice of Patton \cite{patton1990qualitative} in this task. Here, we checked whether the codes within the themes were meaningfully coherent and whether there were clear and identifiable distinctions between the themes. For example, we noticed that the codes included in our initial theme \textit{Technology} were diverse and not coherently connected. Therefore, we broke down this theme into \textit{Tools} and \textit{Infrastructure}. Also, in deciding the naming of the main themes, we considered other available and known DevOps mnemonics (or \textit{models}). This strategy is expected to enable a reader to compare our DevSecOps findings with other studies in DevOps.

We also thematically mapped the challenges to the proposed solutions. In doing the mapping, firstly, we considered how the proposed solution addressed specific challenges, as reported by the paper. Next, we used our own interpretations to map further and update the diagram. Finally, we determined the gap areas by holistically evaluating the currently available solutions against the challenges. The devised gap areas were mapped to the challenges and included in the mapping diagram as well. 

The summary of the method followed in our study is presented in Figure \ref{fig:method}.

\begin{figure*} [t]
    \centering
    \subfloat[\centering Number of papers based on authorship type]{{\includegraphics[width=.45\textwidth]{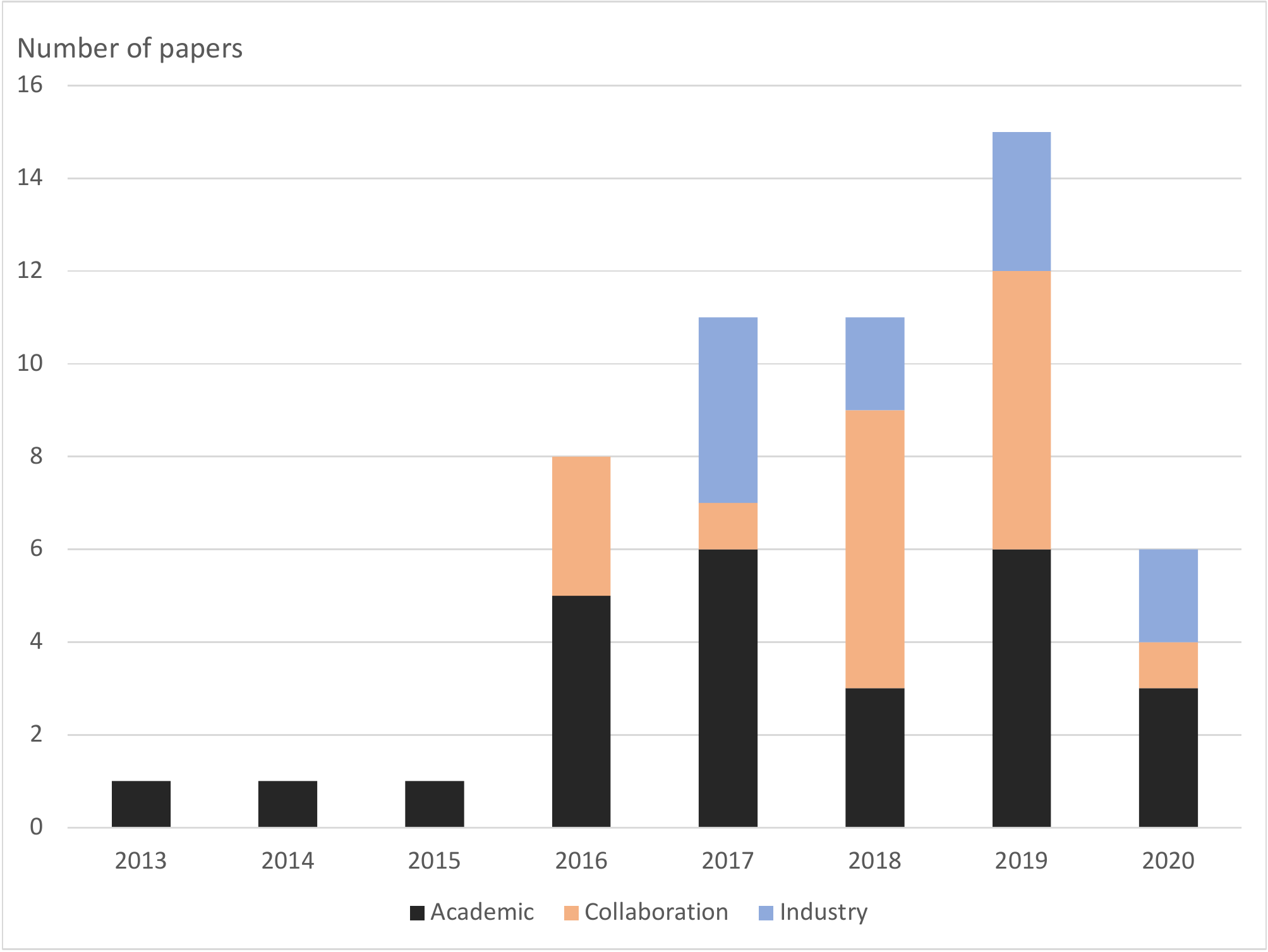} }}%
    \qquad
    \subfloat[\centering Number of papers based on venue type]{{\includegraphics[width=.45\textwidth]{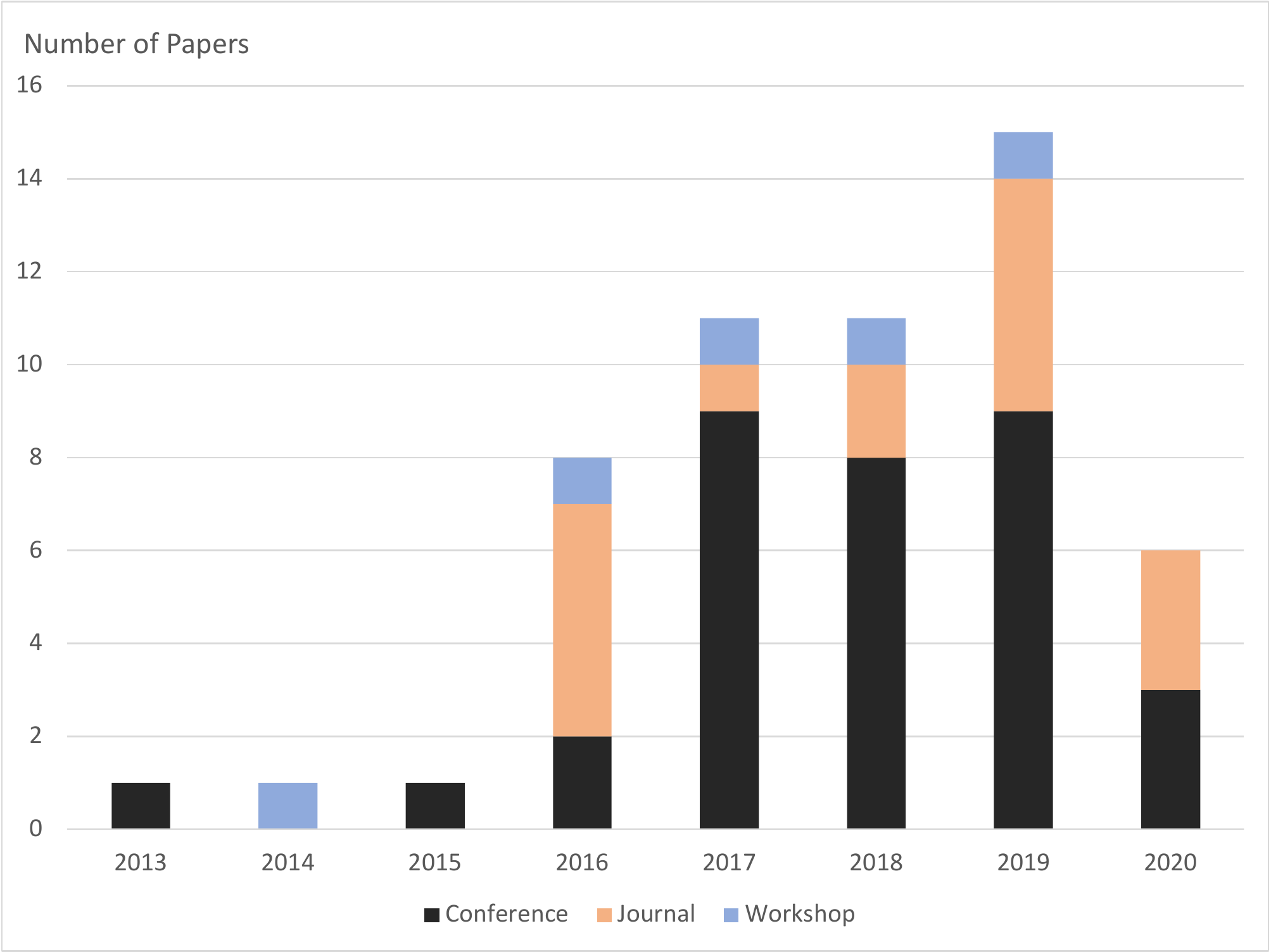} }}%
    \caption{Number of papers included in the review across years}%
\label{fig:trendcharts}    
\end{figure*}

\begin{figure*} [t]
    \includegraphics[width=1\textwidth]{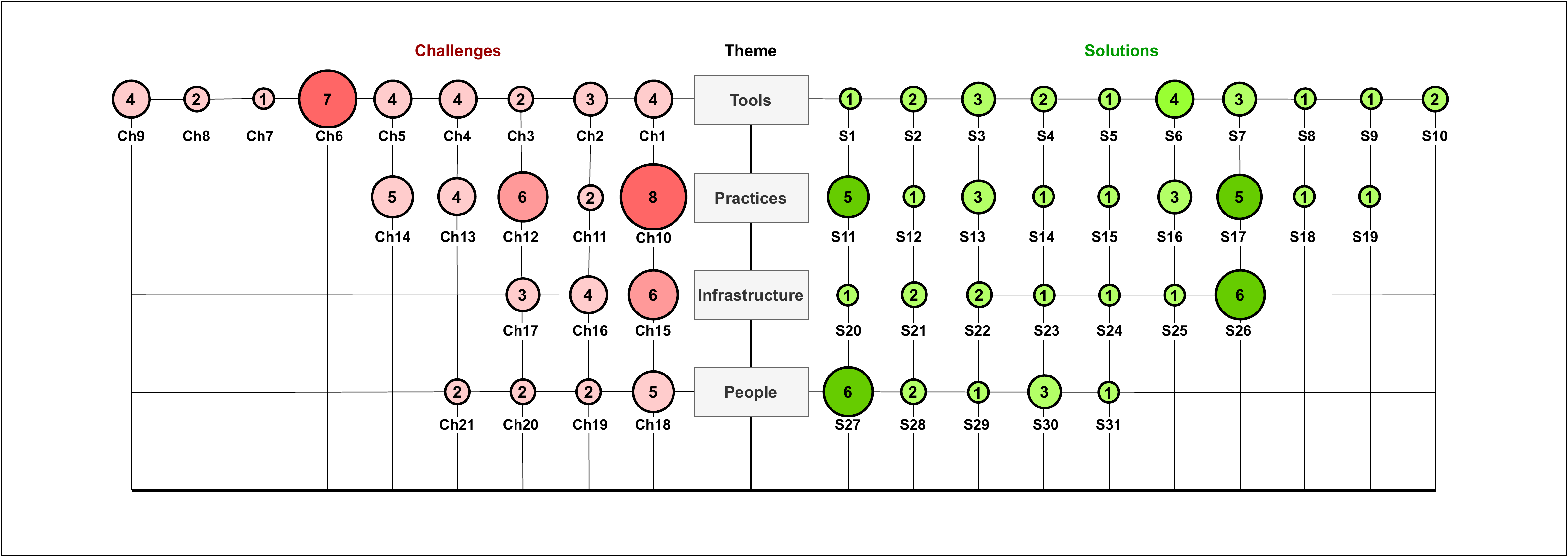}
    \caption{Number of papers per challenge and solution devised in our study: Bubble size based on the number of papers per challenge or solution}
    \label{fig:bubble}
\end{figure*}

\section{Results and Discussion}

In this section, we present the results of our study. First, we present an overview of the primary studies and then discuss the answers to our research questions.

\subsection{Overview of the primary studies}
The reviewed primary studies were published from 2011 to June 2020. Based on these papers, the academic interest for security in the DevOps and CSE domains has emerged in the 2013-2014 period, and there is a gradual upwards growth (Figure 4a). It is also evident that conducting research in this domain is gaining interest in the industry, based on the steadily increasing contributions to the literature from industry-based authors. Figure 4b. depicts the number of papers based on the venue type. Over 60\% of the selected papers (33 out of 54) were published in a large range of conferences (i.e., the work was not limited to specific conferences).

\subsection{Overview of the results}

We categorized our results into the following four main themes.

\begin{itemize}

\item \textbf{People}: This theme covers issues related to knowledge and skills, the collaboration of multidisciplinary team members of DevSecOps, and the organizational culture (e.g., inter-team collaboration issues leading to DevSecOps adoption challenges).

\item \textbf{Practices}: This theme covers issues related to DevOps or continuous practices (e.g., CI/CD) and integrating security practices (e.g., difficulties in integrating manual security practices into DevSecOps)

\item \textbf{Tools}: This theme covers issues related to tools utilized in DevSecOps, their usage scenarios, and the pipeline (e.g., vulnerabilities in containers).

\item \textbf{Infrastructure}: This theme covers issues related to adopting DevSecOps in various types of infrastructures (e.g., complex cloud environments)\newline      
\end{itemize}

In total, we identified 21 challenges and 31 proposed solutions that were classified into the most relevant theme. Figure \ref{fig:bubble} presents these results based on the themes and the number of primary studies per challenge or solution. We also provide separate tables for the devised challenges (Table \ref{tab:challenges}) and solutions (Table \ref{tab:solutions}) with the relevant key points that were used to determine the theme.

\subsection{RQ1: What are the specific DevSecOps adoption challenges?}

This section presents the results for the first research question of our study, categorized under the four themes.

\subsubsection{Challenges related to Tools}
This theme reports the challenges related to using tools in a DevSecOps setting.

\paragraph{Ch1: Challenges related to tool selection}
The usage of tools is highly encouraged in the DevSecOps paradigm. In addition, the DevOps related practices heavily rely on tools \cite{zhu2016devops}. Therefore, there are many tools developed for all stages in DevSecOps \cite{digital.ai}. However, a significant barrier in implementing security into this paradigm is the differences in tool-sets between security and other teams [P52]. Each team member has their own preferences in tools based on specific advantages. This has led to different teams and even the same team members choosing various tool-sets [P04, P52]. 

The empirical studies where developers were interviewed reported that a reason for difficulties in security automation in DevSecOps results from the lack of tool standards [P04]. One reason for this issue is that there is no standard way to perform security automation with a large range of tools [P04, P52].

Further, changing security tools after the project has progressed could lead to adverse effects or conflicts [P28]. For example, Soenen et al. [P28] described how adding a security scanning tool exponentially increased the duration of the integration cycle of their work. Also, the usage of unsuitable deployment tools could have a negative effect on the security of software [P42]. This could only lead to rapidly deploying insecure software.
 
Based on these studies [P04, P28, P42, P52], we observe that the lack of standards for tool selection is a challenge related to the security automation goal of this domain. The tool-centric nature of DevSecOps and the availability of a substantial number of tools exacerbate this problem.

\begin{table*}[hbt!]
\caption{Challenges related to adopting DevSecOps}
\footnotesize
  \begin{tabular} {p{.04\textwidth} | >{\raggedright}p{.26\textwidth}  p{.6\textwidth} |p{.01\textwidth}}
        \toprule 
        \textbf{Theme} & \textbf{Challenges} & \textbf{Key points [Papers which contributed to the point]}& \#\\ \midrule 

        \textbf{Tools} 
         & \textbf{Ch1:} Challenges related to tool selection &     
        \begin{minipage}[t]{\linewidth}
        \begin{itemize}
                    \item No consensus on tool selection across and within teams [P04, P52]
                    \item Selecting wrong tools leading to conflicts with DevSecOps goals [P28, P42]  
                    \end{itemize} \end{minipage} & 25 
                    \\
                    \cmidrule{2-3} 
                    
          & \textbf{Ch2:} Security issues resulting from tool complexity and integration challenges & 
                    \begin{minipage}[t]{\linewidth}
        \begin{itemize}[nosep, noitemsep,topsep=0pt, after=\strut] 
                    \item Complexities of the DevOps and security tools [P33]
                    \item Limitations in documentation resulting in security issues [P33]
                    \item Lack of integrated testing tools [P45, P52]
                    \end{itemize} \end{minipage} 
   &  \\ \cmidrule{2-3} 
   
    & \textbf{Ch3:} Configuration management issues of tools & 
                    \begin{minipage}[t]{\linewidth}
        \begin{itemize}[nosep, noitemsep,topsep=0pt, after=\strut] 
                    \item Using default configurations of security tools [P46]
                    \item Neglecting best practice for configuring software [P47]
                    \end{itemize} \end{minipage} 
   &  \\  \cmidrule{2-3} 
   
    & \textbf{Ch4:} Limitations of static analysis tools affecting rapid deployment cycles & 
                    \begin{minipage}[t]{\linewidth}
        \begin{itemize}[nosep, noitemsep,topsep=0pt, after=\strut] 
                    \item High number of false positives [P04, P25, P26]
                    \item Lengthy code scanning time and resource consumption [P28]
                    \end{itemize} \end{minipage} 
   &  \\  \cmidrule{2-3} 
   
    & \textbf{Ch5:} Limitations of dynamic analysis tools restricting its usage in DevSecOps & 
                    \begin{minipage}[t]{\linewidth}
        \begin{itemize}[nosep, noitemsep,topsep=0pt, after=\strut] 
                    \item Need to be manually run by developers to find security flaws [P17]
                    \item Typically lengthy time needed to run the tool [P22]
                    \item Software/Code needs to be built, installed and configured [P25].
                    \item Limited scope of testing scenarios [P32]
                    \end{itemize} \end{minipage} 
   &  \\  \cmidrule{2-3} 

     & \textbf{Ch6:} Security limitations or vulnerabilities affecting the container ecosystem & 
                    \begin{minipage}[t]{\linewidth}
        \begin{itemize}[nosep, noitemsep,topsep=0pt, after=\strut] 
                    \item Vulnerabilities in containers or container images [P01, P22, P23, P30, P31, P40, P44]
                    \item Security issues due to embedding third party elements \& external intermediaries [P23, P44]
                    \item Insecure configurations and access control settings [P44]
                    \item Casting containers as virtual machines [P23]
                    \end{itemize} \end{minipage} 
   &  \\\cmidrule{2-3} 
     
     & \textbf{Ch7:} Vulnerabilities affecting CI systems & 
                    \begin{minipage}[t]{\linewidth}
        \begin{itemize}[nosep, noitemsep,topsep=0pt, after=\strut] 
                    \item Tenants executing their own code on the CI environment [P34]
                    \end{itemize} \end{minipage} 
   &  \\\cmidrule{2-3} 
   
   & \textbf{Ch8:} Limitations of Infrastructure or Configuration as Code tools and scripts & 
                    \begin{minipage}[t]{\linewidth}
        \begin{itemize}[nosep, noitemsep,topsep=0pt, after=\strut] 
                    \item Security smells of Infrastructure as Code scripts [P43]
                    \item Security issues of Configuration as Code tools [P51]
                    \end{itemize} \end{minipage} 
   &  \\  \cmidrule{2-3} 
     
     & \textbf{Ch9:} Security limitations or vulnerabilities affecting the CD pipeline & 
                    \begin{minipage}[t]{\linewidth}
        \begin{itemize}[nosep, noitemsep,topsep=0pt, after=\strut] 
\item The pipeline opens up additional attack surfaces [P33, P36]
\item Potential security damages related to compromised or misconfigured CDPs [P09]
\item Different team members having same level of access to the pipeline [P09]
\item Security vulnerabilities of the CD pipeline [P36, P50]
    \end{itemize}
    
   \end{minipage} & \\\midrule

   \textbf{Pract.}
         & \textbf{Ch10:} Inability to fully automate traditionally manual security practices to integrate into DevSecOps &
        \begin{minipage}[t]{\linewidth}
        \begin{itemize}[nosep, noitemsep,topsep=0pt, after=\strut] 
                      
                      \item Compliance (testing) practices [P08, P39, P41, P47] 
                      \item Security and privacy by-design practices [P02, P48]
                      \item Architectural risk analysis [P05]
                      \item Risk management practices [P02]
                      \item Threat modelling [P05, P21]
                      \end{itemize}
                      \end{minipage}& 20 \\ \cmidrule{2-3}
                     
        & \textbf{Ch11:} Inability to carry out rapid security requirements assessment &
        \begin{minipage}[t]{\linewidth}
        \begin{itemize}[nosep, after=\strut]
        \item Security requirements assessment not being done before shipping to production  [P06]
        \item Lack of methods to continuously and rapidly assess security requirements [P35]
        \end{itemize}
        \label{tab:challenges}
        \end{minipage}
         &  \\ \cmidrule{2-3}
         
        & \textbf{Ch12:} Challenges related to security measurement practices in rapid deployment environments &
        \begin{minipage}[t]{\linewidth}
        \begin{itemize}[nosep, after=\strut]
        \item Lack of suitable security metrics [P03, P04, P20, P32]
        \item Usage of traditional data or feedback gathering methods [P15, P52]
        \end{itemize}
        \label{tab:challenges}
        \end{minipage}
         &  \\ \cmidrule{2-3}
         
          & \textbf{Ch13:} Challenges related to continuous security assessment &
        \begin{minipage}[t]{\linewidth}
        \begin{itemize}[nosep, after=\strut]
        \item Continuous vulnerability assessment not done in practice [P12, P17, P21]
        \item No consensus on how security measures should be added to the pipeline [P16]
        \end{itemize}
        \label{tab:challenges}
        \end{minipage}
         &  \\ \cmidrule{2-3}
         
          & \textbf{Ch14:} Incompatibility between security and DevOps practices due to velocity of change, complexities and dependencies &
        \begin{minipage}[t]{\linewidth}
        \begin{itemize}[nosep, after=\strut]
        \item Reluctance to adopt DevSecOps due to the perceived incompatibility of security and DevOps practices [P06]
        \item Rapid releases not conducive to thorough testing schemes [P04, P20] 
        \item Developers face trade-offs between speed and security in continuous practices [P49]
        \item Security assurance challenging due to the velocity of change [P52]
           \end{itemize}
    
   \end{minipage} & \\\midrule
   
   \textbf{Infra.} & \textbf{Ch15:} Difficult to adopt DevSecOps in complex cloud environments &
        \begin{minipage}[t]{\linewidth}
        \begin{itemize}[nosep, after=\strut]
        \item Multi-cloud environments [P37, P39, P54]
        \item Systems-of-Systems environments [P03]
        \item Data security in the cloud environment [P18, P24]
        \end{itemize}
        \end{minipage}
         & 13 \\ \cmidrule{2-3} 
         
        & \textbf{Ch16:} Difficult to adopt DevSecOps in resource constrained environments &
        \begin{minipage}[t]{\linewidth}
        \begin{itemize}[nosep, after=\strut]
        \item Embedded systems [P11]
        \item Internet of Things (IoT) systems [P10, P19, P38]
            \end{itemize}
    
   \end{minipage} & \\ \cmidrule{2-3}

         & \textbf{Ch17:} Difficult to adopt DevSecOps in highly regulated environments &
        \begin{minipage}[t]{\linewidth}
        \begin{itemize}[nosep, after=\strut]
        \item Air-gapped environments [P07, P27]
        \item Medical infrastructure [P13]
        \end{itemize}
        \end{minipage}
         &  \\\midrule
         
\textbf{People} & \textbf{Ch18:} Inter-team collaboration issues  &
        \begin{minipage}[t]{\linewidth}
        \begin{itemize}[nosep, noitemsep,topsep=0pt, after=\strut]
            \item Conflicts between developer and security teams [P04, P45]
            \item The need for tools that facilitate collaborative efforts [P53]
            \item Silos between teams [P04, P41, P52]
        \end{itemize}
        \end{minipage} & 09 \\ \cmidrule{2-3}
         
         & \textbf{Ch19:} Knowledge gap in security &
        \begin{minipage}[t]{\linewidth}
        \begin{itemize}[nosep, noitemsep,topsep=0pt, after=\strut]
            \item Lack of security education and training [P04]
            \item Developers lacking security skills [P33]
        \end{itemize}
        \end{minipage} &  \\ \cmidrule{2-3}
        
        & \textbf{Ch20:} Challenges in organizational culture &
        \begin{minipage}[t]{\linewidth}
        \begin{itemize}[nosep, noitemsep,topsep=0pt, after=\strut]
            \item Fear of change or being replaced due to the required cultural/behavioural changes [P03]
            \item Reluctance to prioritize security among team members [P04]
        \end{itemize}
        \end{minipage} &  \\ \cmidrule{2-3}

        & \textbf{Ch21:} Insider threats &
        \begin{minipage}[t]{\linewidth}
        \begin{itemize}[nosep, noitemsep,topsep=0pt, after=\strut]
            \item Misbehaviors due to extensive access [P29, P42]
       \end{itemize}
    
   \end{minipage} & \\\bottomrule

\end{tabular}
\label{table:challenges}
\end{table*}
\normalsize

\paragraph{Ch2: Security issues resulting from tool complexity and integration challenges}
 
Current DevOps and security tools suffer from complexity, particularly related to developers without security training or skills [P33]. The lack of clear documentation for such tools has increased this problem. Studies state that current documentation does not give sufficient information about the security settings of tools [P33]. For example, tool documentation often does not provide details about the \textit{least privilege} security settings. This results in difficulties in setting up tools with the recommended security settings or policies. 

Tool integration (e.g., to form a pipeline) is an essential task in DevSecOps. However, developers are finding it difficult to integrate testing tools into the DevOps pipeline [P45]. This is due to the reason that integrating tools can be a difficult, manual, and time-consuming task [P52].

Based on the above details, a developer is required to have in-depth expertise (\textit{guru-level knowledge} as stated in [P33]) on setting up these tools with the correct security settings or policies and securely integrating them.

\paragraph{Ch3: Configuration management issues of tools}

Configuration management is another problem related to tool usage in DevSecOps. Developers cause vulnerabilities by neglecting the best practice for configuring software and underlying infrastructure [P47].  For example, using default configurations of security tools could lead to wasting resources and producing unsuitable (either excessive or insufficient) security levels for applications [P46]. 

\paragraph{Ch4: Limitations of static analysis tools affecting rapid deployment cycles}

Static Application Security Testing (SAST) (i.e., Static analysis) tools inspect the source, byte, or binary code without running the software [P25]. These tools play an essential role in the early detection of potential faults, vulnerabilities, and code smells [P26].

However, one major problem related to SAST is the considerable time needed to manually assess the substantial number of false positives these tools generate [P04, P25, P26]. In a DevSecOps setting with multiple releases in a short period, carrying out this task can cause delays. Further, teams would also need developers who are knowledgeable in recognizing the false positives [P04].  

The lengthy code scanning time and high resource consumption of SAST tools are also drawbacks for DevSecOps [P28]. In this paradigm, developers are encouraged to commit small amounts of work frequently. However, scanning each incremental work item might not be practical due to the significantly long code scanning times of these tools (especially when the whole codebase is required to be scanned) [P28].

Therefore, based on the above issues, developers find it hard to use static analysis tools in the fast-paced DevSecOps environment [P04].

\paragraph{Ch5: Limitations of dynamic analysis tools restricting its usage in DevSecOps}

Dynamic analysis or DAST tools have many benefits in identifying a wide range of vulnerabilities (e.g., memory safety and input sensitization errors) [P17]. However, software or code must be run to conduct dynamic analysis. This requires software to be built, installed, and configured [P25]. In a DevSecOps setting, where the code is released frequently, conducting these steps at each release is difficult. Another problem with these tools is the fair amount of manual effort needed to set up and run these tools [P17]. Similar to SAST tools, dynamic analysis tools typically take a longer time period to run [P22]. These tools can also be limited in relation to the scope of testing scenarios, which depends on the type of tool being used [P32]. 

All of the above are issues that could hinder the speed and frequency of releases in DevOps. Therefore, despite the abilities of dynamic analysis tools in identifying security defects and vulnerabilities, the drawbacks limit its use in DevSecOps.

\paragraph{Ch6: Security limitations or vulnerabilities affecting the container ecosystem}
Containers are widely used in DevOps. However, vulnerabilities affecting containers and their usage scenarios are widely reported challenges in this domain [P01, P22, P23, P30, P31, P40, P44]. The DevOps community provides a large range of reusable artifacts such as container images \cite{wettinger2017collaborative}. However, these images may be corrupted by attackers and contain vulnerabilities. Although there have been efforts to encourage security assessment of these images from the users, they are often overlooked [P22]. 

Containers embedding various service providers' third-party elements have also increased the number of vulnerabilities [P44]. The increase of external entities which provide code that ends up in the production environment expands the attack surface [P23]. Therefore, developers are faced with the challenge of limiting the reuse of already existing components (e.g., libraries), despite the requirement of fast deployments in DevSecOps.

The practitioners' inappropriate usage of containers has resulted in various other security implications as well [P44]. For example, insecure configurations and access control settings applied by developers can lead to security breaches that can affect all source files stored in the container [P44]. Another example is casting containers as virtual machines by developers, which results in many vulnerabilities [P23]. In this situation, the container is used for requirements for which it was not designed (e.g., embedding more software than supported by the container design), thus increasing the attack surface [P23]).

Based on the above details, despite the advantages of the container ecosystem for DevSecOps, it has also given rise to many security challenges.

\paragraph{Ch7: Vulnerabilities affecting CI systems}
Continuous integration plays a vital role in the DevOps pipeline. To technically enable the CI practice, CI tools have been developed. However, studies report that CI tools are more vulnerable (compared with other tools) for security attacks as tenants execute their own code in the CI environment [P34]. Therefore, there is a higher number of attack vectors related to these systems [P34]. This is a significant security challenge affecting DevSecOps due to the substantial usage of CI systems in this paradigm.

\paragraph{Ch8: Limitations of Infrastructure or Configuration as Code tools and scripts}
Infrastructure as Code (IaC) tools are highly utilized in DevSecOps, due to their use in configuring infrastructure rapidly. However, critical security challenges related to the scripts used in the tools have been reported. For example, studies have discovered that practitioners mistakenly introduce many security smells into IaC scripts [P43]. Here, security smells are indicative of security weaknesses and can lead to security breaches [P43]. 

Configuration as Code (CaC) tools are another type of tool used heavily in CD, which enables the management of computing and network configurations through source code \cite{humble2010continuous}. However, security-related issues in CaC tools are not given priority in the community [P51]. This was evident by a study that mined Stack Overflow posts to identify developer challenges in using CaC tools [P51]. As one of the findings, the authors noted that despite security-related questions of such tools being common, they often do not get satisfactory answers from practitioners.

These results point towards developers' lack of interest in focusing on security issues of IaC and CaC tools which is problematic as any related faults could lead to substantial damages. 

\paragraph{Ch9: Security limitations or vulnerabilities affecting the CD pipeline}
A critical security problem related to CD of software is the security limitations or vulnerabilities of the CD pipeline (CDP) itself. A typical CDP is not designed in a manner that gives security requirements much prominence [P09]. Further, a number of separate tool suites and users are involved in different CDP stages. Most of the technical components of the CDP run in an environment with several online interfaces, and these components are vulnerable to various kinds of malicious attacks [P36]. Therefore, the CDP opens up additional attack surfaces which can be exploited [P33]. Studies note that a compromised or misconfigured CDP may result in malicious code or unwanted debugging/experimental code ending up in the production environment [P09]. This is critical as malicious software with customer access can lead to adverse consequences. 

DevSecOps advocates for a high level of team-collaboration. However, different team members (Dev/Sec/Ops) having the same level of access to the pipeline can lead to security-related challenges [P09]. Paule et al. [P50] assessed two CDPs from industry projects using a threat modeling approach. They state that even though most team members have access to CDP configurations, these members pose a risk to the infrastructure and the application due to the lack of security knowledge and awareness. Therefore, any potential damages could be intentional or unintentional. This study also reported on other vulnerabilities such as unencrypted connections and insecure (e.g., foreign or customer) environments affecting the CDP [P50]. 
\subsubsection{Summary of the challenges related to Tools}

\begin{tcolorbox}[left=1pt, top=1pt, right=1pt, bottom=1pt]

\begin{itemize}
    \item Developers are finding it difficult to select or use the increasing number of security tools due to the lack of standards, documentation, and training. The complexity, integration, and configuration challenges of such tools aggravate this problem [Ch1, Ch2, Ch3].

\end{itemize}

\end{tcolorbox}

\begin{tcolorbox}[left=1pt, top=1pt, right=1pt, bottom=1pt]

\begin{itemize}
    \item The inability of established security tools to support the rapid deployment of software is a significant challenge. Therefore, there is a tendency among practitioners to not utilize these tools despite their benefits [Ch4, Ch5].
    \item A large number of security vulnerabilities are affecting popular tools used in the DevSecOps pipeline (e.g., containers, CI systems) and the pipeline itself [Ch6-Ch9].
\end{itemize}

\end{tcolorbox}

\subsubsection{Challenges related to Practices}
This theme reports the challenges related to conducting certain practices (e.g., security practices) in a DevSecOps setting.

\paragraph{Ch10: Inability to fully automate traditionally manual security practices to integrate into DevSecOps}

In DevOps, automation plays a significant role due to the requirement of rapid and continuous releases. To achieve this goal, DevOps contains a set of continuous practices (e.g., CI/CDE/CD), which are automated to a large extent. However, automation of security practices has become problematic as many of them are traditionally performed manually. Our study captured several such practices. Many studies have stated that conducting compliance practices has become challenging with the pace of DevOps [P08, P39, P41, P47]. In this scenario, compliance of standards, frameworks, and best practice processes and the velocity of DevOps practices act as opposing factors [P08]. 

Other security practices which are challenging to integrate into DevOps include: security or privacy by design [P02, P48], architectural risk analysis [P05], threat modeling [P05, P21], and risk management [P02]. The key reason for this difficulty is the need for substantial human input to execute these processes. As a result, they can be time-consuming and would have a negative impact on rapid releases.

\paragraph{Ch11: Inability to carry out rapid security requirements assessment}
The process of conducting rapid assessment of security requirements is seen as difficult in a DevSecOps setting. Due to the fast pace of CDs, it is challenging to thoroughly verify the security requirements before shipping software to a production environment. Therefore, such an assessment of security requirements is not carried out in a practical setting [P06].  Lack of tools and methods to carry out this process could be a key reason for this situation [P35].

\paragraph{Ch12: Challenges related to security measurement practices in rapid deployment environments}
The task of measuring security in software is hugely challenging \cite{jaatun2012hunting}. Measuring security in the DevOps paradigm is even more difficult due to the continuous and rapid software releases [P20]. Lack of suitable security metrics [P03, P04, P20, P32] was a frequently cited reason for these difficulties.

Another challenge related to security measurement in rapid deployment environments was the usage of traditional (and slow) data gathering methods. Fast feedback loops are required for CD systems [P15]. Therefore, the usage of traditional methods hinders this task. 

Quick feedback loops between the teams in DevSecOps and other relevant project stakeholders are also highlighted as an important requirement [P52]. This is important in terms of maintaining traceability to enable fault localization and resolving issues. However, these processes are difficult to implement in practice due to problems ranging from using traditional methods [P15] to cultural issues.

\paragraph{Ch13: Challenges related to continuous security assessment}
Continuous security [P14] assessment is a recommended practice in DevSecOps. However, processes related to this practice are not widely adopted. Continuous vulnerability assessment is one such process [P12, P17, P21]. This is due to practitioners not carrying out periodic checks for vulnerabilities [P12], team members lacking knowledge of continuous vulnerability assessment [P21], and other scaling issues related to continuous security testing [P17].

For continuous security assessment to be a success, there must be clear instructions on which sections of the pipeline security measures will be included. However, there is a lack of consensus (i.e., a standardized methodology) on how security measures must be included in a DevOps pipeline [P16]. 

\paragraph{Ch14: Incompatibility between security and DevOps practices due to velocity of change, complexities, and dependencies}
While one of the key aims of DevOps is the speed of release, many security testing practices require human input. For example, penetration testing requires substantial human input (even though there are tools for penetration testing, human input is required to configure, run and then assess outputs) [P04]. Therefore, these are time-consuming practices. As a result, rapid releases are not conducive for thorough testing programs [P20]. In addition, the increasing complexities, vulnerabilities, and dependencies to third-party components such as open-sourced libraries have made security assurance even more challenging [P52]. Due to these reasons, developers face trade-offs between the speed of release and security [P49], and organizations see maintaining the velocity of DevOps and thorough security assurance as incompatible practices [P06].

\subsubsection{Summary of the challenges related to Practices}

\begin{tcolorbox}[left=1pt, top=1pt, right=1pt, bottom=1pt]

\begin{itemize}
    \item The inability to automate traditionally manual security practices to fit into the DevSecOps paradigm is a critical challenge yet to be adequately addressed [Ch10, Ch11].
    \item DevOps practices focus on speed and agility. The key focus of security practices is security assurance through comprehensive testing, which is time-consuming. Ultimately, developers face trade-offs between speed and security in a DevOps setting [Ch10-Ch14]
    \item Due to these reasons, some organizations are reluctant to transform into DevOps due to this perceived incompatibility between security and DevOps [Ch14].
\end{itemize}
\end{tcolorbox}

\subsubsection{Challenges related to Infrastructure}
In this section, we present the challenges related to adopting DevSecOps in certain types of infrastructures.

\paragraph{Ch15: Difficult to adopt DevSecOps in complex cloud environments}

DevSecOps principles and practices are difficult to be adopted in various types of complex cloud environments. For example, producing secure software rapidly is challenging in a cloud environment when the target system is in the form of system-of-systems (SoS) [P03]. SoS are complex systems that are comprised of other constituent systems [P03]. 

Another type of challenging infrastructure is multi-cloud environments (applications that combine multiple heterogeneous cloud offerings) [P37, P39]. Security assurance has become challenging in architectures such as microservices [P54] and automated distributed deployments [P37] which are heavily utilized in multi-cloud environments. Studies also report that data security is another critical issue in this domain.  Producing software rapidly while ensuring data security in such distributed and heterogeneous complex cloud environments is a complex task [P18, P24].

\paragraph{Ch16: Difficult to adopt DevSecOps in resource-constrained environments}
We captured two types of resource-constrained environments where adopting DevSecOps was reported to be challenging: Internet of Things (IoT) systems and embedded systems. Here, the challenges related to IoT include high heterogeneity of the infrastructure (which increases the attack surface) and the complexity in maintaining and evolving such systems [P10]. Also, due to the distributed and heterogeneous nature of IoT-based systems, setting up secure deployment pipelines and monitoring security-related events are challenging [P38]. It has also been reported that there is a lack of key enabling tools that could result in the low adoption of secure DevOps practices in IoT [P19].

Regarding embedded systems, M{\aa}rtensson et al. [P11] present a set of factors that must be considered when applying CI into software-intensive embedded systems. Firstly, they state that compliance with standards, which is critical in embedded systems, changes the focus away from delivering working software rapidly. Secondly, another factor is the restriction for information access resulting from security concerns in the domain. Therefore, such factors would hinder the implementation of DevSecOps practices in these types of infrastructures.

\paragraph{Ch17: Difficult to adopt DevSecOps in highly regulated environments}
Highly regulated environments were the next type of complex infrastructure that was captured from this theme. The studies showed how secure continuous practices were challenging in patch delivery for medical devices [P13] and air-gapped production environments [P07]. Characteristics of regulated environments (as reported in P07 and P27) such as zero-trust security architectures, segregated environments, temporary access policies, and restricted communication with stakeholders make it challenging to implement secure DevOps practices. Further, if the regulated environment or infrastructure has an \textit{air-gap} (i.e., no direct production access, or many security gates) to production environments, adoption of certain continuous practices can be challenging.

\subsubsection{Summary of the infrastructure-related challenges}

\begin{tcolorbox}[left=1pt, top=1pt, right=1pt, bottom=1pt]

\begin{itemize}
    \item Most of the problems in this theme arose due to the nature of certain challenging infrastructures (e.g., distributed, heterogeneous or segregated environments) or restrictive policies (e.g., access control) conflicting with DevSecOps principles or practices [Ch15, Ch16, Ch17].
\end{itemize}
\end{tcolorbox}

\subsubsection{Challenges related to People}
This section presents the reported challenges related to people. 

\paragraph{Ch18: Inter-team collaboration issues}
Strong communication and collaboration across teams are key success factors in DevSecOps. However, the most number of papers categorized under the theme \textit{People} were challenges related to \textit{inter-team collaboration}. One of the main collaboration issues reported was the conflicts between development and security teams [P04, P45]. Studies reported developer sentiments such as how they feel like security team members judge and criticize the work done by them [P04]. Further, developers were unhappy about losing the autonomy of their own development work [P04]. 

The silo-based work culture in the software community [P04] is a barrier to secure DevOps [P52]. These silos hinder frequent and effective communication and collaboration between stakeholders. With regard to secure DevOps, security team members are important stakeholders who need to be part of a project. The other teams are required to communicate frequently and collaborate with security team members, and have an attitude of shared responsibility [P04]. However, studies captured by our SLR showed that this was not the case in practice [P41]. There is also a need for tool-supported automation to facilitate collaborative efforts with the security team [P53]. For example, in the early stages of the development process, tool support would be helpful to manage and provision design models and understand the design rationales of the security team or other teams members in a collaborative manner [P53].

\paragraph{Ch19: Knowledge gap in security}
DevSecOps advocates developers to engage in security tasks. However, the lack of security skills and knowledge of developers hinders this goal. Developers lacking security education and training contribute to this problem. Studies state that one reason for this situation is software engineering and software security education being separate [P04].

Wilde et al. [P33] offer a different point of view. They argue that it is difficult to have a sufficient number of personnel with the required security skills. Therefore, the security aspects of DevOps should be simplified so that those tasks can be managed by developers who lack specialized security skills [P33]. 

\paragraph{Ch20: Challenges in organizational culture}

For successful DevSecOps adoption, many cultural and behavioral changes are required. However, this has resulted in competition and fear among people [P03]. For example, fear of being replaced or no longer being recognized and resentful sentiments of not being the owner of the foreground are some issues reported [P03].

Reluctance to prioritize security is another reported challenge related to organizational culture [P04]. In this case, authors have described instances where developers or other team members not fully taking responsibility for security practices (e.g., P04 quotes a developer: \textit{\say{nobody wants to take responsibility for security because it adds nothing}}). They also share that in certain companies, the management does not prioritize security [P04]. 

\paragraph{Ch21: Insider threats}

The role of developers has changed in DevOps, where now they have access to production services.  Therefore, a larger number of insiders have access to the production environment [P29]. Consequently, if any one of these insiders turns rogue, they can cause harm via access to the production setting or governance tools used to control that environment [P29]. Therefore, unrestricted collaboration might lead to inappropriate access to the system resources [P42]. Based on the above, we can see that the risk of insider threats has increased with the change of roles in DevSecOps.

\subsubsection{Summary of the people related challenges}

\begin{tcolorbox}[left=1pt, top=1pt, right=1pt, bottom=1pt]
\begin{itemize}
    \item Challenges reported in this theme mainly resulted from the inability of Dev/Sec/Ops team members or the management to engage in the required culture change of this paradigm [Ch18, Ch20]. 
    \item Developers lacking security skills is a critical issue in this domain, as developers are required to carry out certain security practices in DevSecOps [Ch19]. 
\end{itemize}
\end{tcolorbox}

\begin{table*}[t!]
\centering
\caption{Solutions proposed to address DevSecOps challenges}
\footnotesize
  \begin{tabular} { p{.04\textwidth} | >{\raggedright}p{.25\textwidth}  p{.6\textwidth} |p{.02\textwidth}}
        \toprule 
        \textbf{Main theme} & \textbf{Solutions} & \textbf{Key points [Papers which contributed to the point]}& \#\\ \midrule 

        \textbf{Tools} &
                    \textbf{S1:} Practitioners converge towards tool standards &     
                    \begin{minipage}[t]{\linewidth}
                    \begin{itemize}[nosep, noitemsep,topsep=0pt, after=\strut] 
                    \item The community converging towards standards in tool selection and usage [P04]
                    \end{itemize} \end{minipage} 
                    & 19 \\\cmidrule{2-3}
                    
                    & \textbf{S2:} Documentation with security support & 
                    \begin{minipage}[t]{\linewidth}
                    \begin{itemize}[nosep, noitemsep,topsep=0pt, after=\strut] 
                    \item Clear documentation detailing security settings [P43]
                    \item Documentation delivered through centralized repositories [P06]
                    \end{itemize} \end{minipage} 
                    &  \\ \cmidrule{2-3}
                    
                    & \textbf{S3:} Adopting best practice for tool usage &
                    \begin{minipage}[t]{\linewidth}
                    \begin{itemize}[nosep, noitemsep,topsep=0pt, after=\strut] 
                    \item  Tools configured based on the context, without using default settings [P46]
                    \item Being specific on the nature of the testing (e.g., excluding non production code) [P26]
                    \item  Setting up optimized tool pipelines [P28]
                    \end{itemize} \end{minipage} 
                    &  \\ \cmidrule{2-3}
                    
                    & \textbf{S4:} Move to cloud-based solutions & 
                    \begin{minipage}[t]{\linewidth}
                    \begin{itemize}[nosep, noitemsep,topsep=0pt, after=\strut] 
                    \item  Using static code analysis as a service [P25]
                    \item Cloud tools to facilitate collaboration across cross-disciplinary teams [P53]
                    \end{itemize} \end{minipage} 
                    &  \\ \cmidrule{2-3}
                    
                    & \textbf{S5:} Interactive application security testing (IAST) tools &
                    \begin{minipage}[t]{\linewidth}
                    \begin{itemize}[nosep, noitemsep,topsep=0pt, after=\strut] 
                    \item  Using IAST tools which combine the SAST and DAST tool features  [P32]
                    \end{itemize} \end{minipage} 
                    &  \\ \cmidrule{2-3}
                    
                    & \textbf{S6:} Tools for continuous vulnerability assessment &
                    \begin{minipage}[t]{\linewidth}
                    \begin{itemize}[nosep, noitemsep,topsep=0pt, after=\strut]
                    \item Enabling continuous vulnerability assessment [P17, P40]
                    \item Combining multiple static and dynamic analysis tools [P22, P31]
                    \end{itemize} \end{minipage} 
                    &  \\ \cmidrule{2-3}
                    
                    & \textbf{S7:} Using orchestration platforms &
                    \begin{minipage}[t]{\linewidth}
                    \begin{itemize}[nosep, noitemsep,topsep=0pt, after=\strut] 
                    \item Orchestrators to limit the misuse of containers [P23] 
                    \item To enable better isolation [P44]
                    \item To reduce the container attack surface [P30]
                    \end{itemize} \end{minipage} 
                    &  \\ \cmidrule{2-3}
                    
                    & \textbf{S8:} Using a virtualization tool to encapsulate part of the system &
                    \begin{minipage}[t]{\linewidth}
                    \begin{itemize}[nosep, noitemsep,topsep=0pt, after=\strut] 
                    \item Encapsulate the build job to secure the build server [P34]
                    \end{itemize} \end{minipage} 
                    &  \\ \cmidrule{2-3}
                    
                    & \textbf{S9:} Static analysis for IaC Scripts &
                    \begin{minipage}[t]{\linewidth}
                    \begin{itemize}[nosep, noitemsep,topsep=0pt, after=\strut] 
                    \item  To automatically identify security smells in IaC scripts [P43]
                    \end{itemize} \end{minipage} 
                    &  \\ \cmidrule{2-3}
                    
                    & \textbf{S10:} Reusable design fragments and security tactics &
                    \begin{minipage}[t]{\linewidth}
                    \begin{itemize}[nosep, noitemsep,topsep=0pt, after=\strut] 
                    \item  Design fragments to secure the tool pipeline [P09]
                    \item Security tactics to increase the security of a CD pipeline [P36]
                    \end{itemize} \end{minipage} & \\\midrule
                    
                    \textbf{Pract.}
                    
                    & \textbf{S11:} Adapting standards, policies, models, service level agreements (SLA) into testable criteria &     
                    \begin{minipage}[t]{\linewidth}
                    \begin{itemize}[nosep, noitemsep,topsep=0pt, after=\strut] 
                    \item SLAs to model and assess security requirements or made machine readable [P02, P37, P39]
                    \item Adapting information security standards into testable criteria [P08]
                    \item Continuous automated compliance testing mechanisms [P47]
                    \end{itemize} \end{minipage} 
                    & 16 \\ \cmidrule{2-3}
                    
                    & \textbf{S12:} Automated vulnerability detection through requirement analysis &     
                    \begin{minipage}[t]{\linewidth}
                    \begin{itemize}[nosep, noitemsep,topsep=0pt, after=\strut] 
                    \item Toolchains that automatically translate requirements documents [P35].
                    \end{itemize} \end{minipage} 
                    &  \\\cmidrule{2-3}
                    
                    & \textbf{S13:} Devising security metrics or metric based approaches &     
                    \begin{minipage}[t]{\linewidth}
                    \begin{itemize}[nosep, noitemsep,topsep=0pt, after=\strut] 
                    \item Security metrics based on developer attributes and activities [P04]
                    \item Measure second-order effects in the development process [P20]
                    \item A taxonomy of metrics for IoT [P38]
                    \end{itemize} \end{minipage} 
                    &  \\\cmidrule{2-3}
                    
                    & \textbf{S14:} Effective process documentation and logging strategies &     
                    \begin{minipage}[t]{\linewidth}
                    \begin{itemize}[nosep, noitemsep,topsep=0pt, after=\strut] 
                    \item Security logging of user processes using specific tools [P06]
                    \item Automated process documentation [P06]
                    \end{itemize} \end{minipage} 
                    &  \\\cmidrule{2-3}
                    
                    & \textbf{S15:} Big data and behavioral analytics techniques &     
                    \begin{minipage}[t]{\linewidth}
                    \begin{itemize}[nosep, noitemsep,topsep=0pt, after=\strut] 
                    \item Obtain fast feedback from the end user of systems [P15]
                    \item Predictive analytics to be aware of trends in user behaviours [P15]
                    \end{itemize} \end{minipage} 
                    &  \\\cmidrule{2-3}
                    
                    & \textbf{S16:} Shifting security to the left &     
                    \begin{minipage}[t]{\linewidth}
                    \begin{itemize}[nosep, noitemsep,topsep=0pt, after=\strut] 
                    \item  Giving a higher priority to security and identifying security issues at a very early stage  [P08]
                    \item Overcome costly fixes at a later stage [P04]
                    \item Having the correct tool sets in place early [P16]
                    \end{itemize} \end{minipage} 
                    &  \\\cmidrule{2-3}
                    
                    & \textbf{S17:} Implementing continuous security assessment practices &     
                    \begin{minipage}[t]{\linewidth}
                    \begin{itemize}[nosep, noitemsep,topsep=0pt, after=\strut] 
                    \item Security treated as a key concern across all stages [P14]
                    \item A smart and lightweight approach to identify security vulnerabilities [P14]
                    \item Consensus required on how security practices would be included in the cycle [P16]
                    \item Required tools need to be in place [P12]
                    \item Continuous monitoring as an example for a specific continuous security practice [P07, P38]
                    \end{itemize} \end{minipage} 
                    &  \\\cmidrule{2-3}
                    
                    & \textbf{S18:} Security patch management using DevOps practices &
                    \begin{minipage}[t]{\linewidth}
                    \begin{itemize}[nosep, noitemsep,topsep=0pt, after=\strut] 
                    \item  Addresses security vulnerabilities in a rapid manner [P20]
                    \end{itemize} \end{minipage} 
                    &  \\ \cmidrule{2-3}
                    
                    & \textbf{S19:} Using threat analysis practices &
                    \begin{minipage}[t]{\linewidth}
                    \begin{itemize}[nosep, noitemsep,topsep=0pt, after=\strut] 
                    \item  Using STRIDE threat modeling approach to identify vulnerabilities in a CDP [P50]  
                       \end{itemize}
   \end{minipage} & \\ \midrule
   
                     \textbf{Infra.}
                    & \textbf{S20:} Strict access management and policies &
                    \begin{minipage}[t]{\linewidth}
                    \begin{itemize}[nosep, noitemsep,topsep=0pt, after=\strut] 
                    \item Need-based-access [P07].
                    \item Changes to production made automated [P07]
                    \end{itemize} \end{minipage} 
                    & 11 \\ \cmidrule{2-3}
                    
                    & \textbf{S21:} Adopting Infrastructure as code &
                    \begin{minipage}[t]{\linewidth}
                    \begin{itemize}[nosep, noitemsep,topsep=0pt, after=\strut] 
                    \item Infrastructure to be the versioned, tested, built and deployed using IaC in Air-gapped environments [P07].
                    \item IaC support for setting up pre-configured systems and networks [P27]
                    \end{itemize} \end{minipage} 
                    &  \\ \cmidrule{2-3}               
                    
                    & \textbf{S22:} Creating simulation or replication environments for testing &
                    \begin{minipage}[t]{\linewidth}
                    \begin{itemize}[nosep, noitemsep,topsep=0pt, after=\strut] 
                    \item To enable internal testing in highly regulated environments [P27].
                    \item To enable testing application scenarios against programmed circumstances [P19]
                    \end{itemize} \end{minipage} 
                    &  \\ \cmidrule{2-3}
                    
                    & \textbf{S23:} Model driven engineering to support DevSecOps &
                    \begin{minipage}[t]{\linewidth}
                    \begin{itemize}[nosep, noitemsep,topsep=0pt, after=\strut] 
                    \item To address the heterogeneity of IoT infrastructures [P10].
                    \end{itemize} \end{minipage} 
                    &  \\ \cmidrule{2-3}

          \multicolumn{1}{r}{} & & \multicolumn{1}{r}{Continues to the next page} & \\ 
  
\end{tabular}
\label{tab:sol_tools}
\end{table*} 

\begin{table*}[t!]
\ContinuedFloat
\centering
\caption{Continued from the previous page}
\footnotesize
  \begin{tabular} { p{.04\textwidth} | >{\raggedright}p{.25\textwidth}  p{.57\textwidth} |p{.02\textwidth}}
        \toprule 
        \textbf{Main theme} & \textbf{Solutions} & \textbf{Key points [Papers which contributed to the point]}& \#\\ \midrule 
                    
                    & \textbf{S24:} Systematic evaluation of product-specific vulnerabilities &
                    \begin{minipage}[t]{\linewidth}
                    \begin{itemize}[nosep, noitemsep,topsep=0pt, after=\strut] 
                    \item Design aware risk assessment in highly regulated industries [P13].
                    \end{itemize} \end{minipage} 
                    &  \\ \cmidrule{2-3}
                    
                    & \textbf{S25:} Hybrid life cycles with data-security focus &
                    \begin{minipage}[t]{\linewidth}
                    \begin{itemize}[nosep, noitemsep,topsep=0pt, after=\strut] 
                    \item Combining data security and software development life cycles [P18].
                    \end{itemize} \label{tab:sol_infra} \end{minipage} 
                    &  \\   \cmidrule{2-3}
                    
                    & \textbf{S26:} Framework support for DevSecOps &
                    \begin{minipage}[t]{\linewidth}
                    \begin{itemize}[nosep, noitemsep,topsep=0pt, after=\strut] 
                    \item Author defined frameworks  [P03, P19, P24, P37, P39, P54]
                        \end{itemize}
    
   \end{minipage} & \\\midrule
   
   \textbf{People}
        & \textbf{S27:} Facilitating inter-team communication and collaboration with the appropriate controls or standards &     
                    \begin{minipage}[t]{\linewidth}
                    \begin{itemize}[nosep, noitemsep,topsep=0pt, after=\strut] 
                    \item Inter-team collaboration as a best practice [P06, P42]
                    \item Forming multidisciplinary teams [P08]
                    \item Short feedback cycles with other teams [P42]
                    \item Increase developer engagement in security tasks [P41]
                    \item Clear separation of duties [P06]
                    \item The need for standardized communication strategies [P13, P27] 
                    \end{itemize} \end{minipage} 
                    & 10 \\\cmidrule{2-3}
                    
                    & \textbf{S28:} Having security champions in teams &
                    \begin{minipage}[t]{\linewidth}
                    \begin{itemize}[nosep, noitemsep,topsep=0pt, after=\strut] 
                    \item Security champions to bridge the gap between security and development [P04]
                    \item To reduce developer resistance [P20]. 
                    \end{itemize} \end{minipage} 
                    &  \\ \cmidrule{2-3}
                    
                    & \textbf{S29:} Carrying out organizational HRM programs in parallel &
                    \begin{minipage}[t]{\linewidth}
                    \begin{itemize}[nosep, noitemsep,topsep=0pt, after=\strut] 
                    \item HRM programs to address challenges in culture change [P03]
                    \end{itemize} \end{minipage} 
                    &  \\ \cmidrule{2-3}
                    
                    & \textbf{S30:} Implementing security knowledge sharing methods and training &
                    \begin{minipage}[t]{\linewidth}
                    \begin{itemize}[nosep, noitemsep,topsep=0pt, after=\strut] 
                    \item To enable developers use security tools and assess the outputs [P04]
                    \item Blameless security retrospectives [P04]
                    \item To identify when to seek advice from the security team [P20]
                    \item Examples of specific security training activities [P42]
                    \end{itemize} \end{minipage} 
                    &  \\ \cmidrule{2-3}

                    & \textbf{S31:} Integrity protection frameworks &
                    \begin{minipage}[t]{\linewidth}
                    \begin{itemize}[nosep, noitemsep,topsep=0pt, after=\strut] 
                    \item Framework for holistic integrity protection in microservice-based systems [P29]
                    \end{itemize} \end{minipage} 
                    & \\\bottomrule

\end{tabular}
\label{tab:solutions}
\end{table*}

\subsection{RQ2: What are the solutions proposed to address the DevSecOps adoption challenges?}
This section presents the results for the second research question, the proposed solutions in DevSecOps.

\subsubsection{Solutions proposed related to tools}
This section reports the solutions proposed for the tool-related challenges.

\paragraph{S1: Practitioners converge towards tool standards}
Studies reported that developers were finding the tool selection challenging due to the high number of tools available for each stage in DevSecOps [P04]. To address this problem, developers were advocating that the community should converge towards tool standards (e.g., selection and usage) [P04]. In doing so, there would be more commonly accepted selection and usage guidelines, which would reduce a large number of tool-related challenges. 

\paragraph{S2: Documentation with security support}
The complexity of the available DevSecOps tools was a reported problem [P33]. The lack of thorough documentation seemed to exacerbate this issue. Therefore, better documentation related to the usage of tools is advocated [P43]. This would be a solution for the configuration management challenges as well as security settings related to tools. The optimal and recommended configurations and security settings for tools can be delivered through clear documentation [P43]. Further, the documentation can be made easily accessible for all team members by using a centralized repository [P06].

\paragraph{S3: Adopting best practice for tool usage} 
Best practice related to tools needs to be followed to ensure quality attributes such as security. For example, keeping the default configuration settings could lead to issues such as unsuitable security (excessive or insufficient) levels for software [P46]. A best practice in this scenario is to configure the tool based on the context (e.g., technology stack) in which it is being used. 

Studies have also offered a range of best practice regarding using scanning tools in the pipeline, for example, being clear on what specific checks to perform on a project (e.g., web specific checks would not make sense for a non-web based project), excluding certain non-production code from scanning, properly configuring tools based on the internal guidelines, and avoiding duplicate checks can be noted [P26]. 

Another example related to using SAST tools is setting up optimized tool pipelines [P28]. Here, studies discussed running static analysis tools in parallel as a solution to the lengthy code scanning time (e.g., particularly for deep scans) [P28].

\paragraph{S4: Move to cloud-based solutions}
Moving to cloud services is a popular trend in the current context. By using such services, customers are able to avoid certain drawbacks of standalone tools. This is true concerning tools that are used in DevSecOps as well [P25]. For example, static analysis as a service solutions were reported [P25]. By using such solutions, users are able to minimize certain DevSecOps adoption challenges related to SAST tools, such as high amounts of false positives and configuration or setup difficulties of these tools. 

A cloud-based service (i.e., CAIRIS) was also recommended to facilitate inter-team collaboration between security and usability teams [P53]. By using this service, security and usability engineers are better able to manage machine-readable design models (i.e., design as code), leading to higher collaboration among their teams [P53].

\paragraph{S5: Interactive application security testing (IAST) tools}
IAST is a testing tool that integrates well with the DevOps paradigm. This is a new class of security testing tools that combines the features of static and dynamic analysis tools [P32]. These tools are typically run during the functional unit testing stage. Both commercial and open-sourced IAST tools that are suited to modern software development methods are available [P32].

\paragraph{S6: Tools for continuous vulnerability assessment}

Tools enabling continuous security assessment practices are a key solution proposed in DevSecOps. For example, a tool that checks for vulnerabilities each time there is a code commit was reported [P17]. Such continuous assessments can alert developers of security issues immediately after they are introduced [P17, P40]. 
Another proposal was to use multiple static and dynamic analysis tools for the vulnerability assessment of containers [P22, P31]. By doing so, drawbacks of individual tools, such as false positives, can be mitigated. 

\paragraph{S7: Using orchestration platforms}
Studies have widely proposed using orchestration platforms as a solution for the challenges related to containers [P23, P30, P44]. Orchestrators offer many advantages to limit the misuse of containers [P23]. For example, orchestrators enable a high level of abstraction (e.g., task/replication controllers) and remote persistent storage, thus allowing for better isolation [P44]. Other security features of such an orchestration platform should include: enabling container image sanity, reducing the container attack surface, tightening user access control, and hardening of the host [P30].

\paragraph{S8: Using a virtualization tool to encapsulate part of the system}
This is a solution proposed to address attacks to the CI or build servers. Here, the virtualization tool should encapsulate the build job resulting in a secure build server. In this proposal, the aim is to set the build server to an un-compromised state after each build job [P34].

\paragraph{S9: Static analysis for IaC scripts}
This is a solution proposed to address security weaknesses (e.g., security smells) of IaC scripts. Such scripts are the input to IaC tools, which are heavily utilized in DevSecOps.  In this case, Rahman et al. [P43] developed a static analysis tool called \textit{Security Linter for Infrastructure as Code scripts (SLIC)} to automatically identify security smells in IaC scripts.

\paragraph{S10: Reusable design fragments and security tactics}
A proposed solution to address the security issues of a CDP is the usage of reusable verified design fragments. These design fragments are introduced to package security patterns and platform features [P09]. Rimba et al. [P09] showed how four primitive tactics could compose design fragments into a secure pipeline.
Ullah et al. [P36] also demonstrated how security tactics could be used to increase the security of a CDP. They focused on the access control for the repository, main server, and CI server. 

\subsubsection{Summary of the tools related solutions}

\begin{tcolorbox}[left=1pt, top=1pt, right=1pt, bottom=1pt]
\begin{itemize}
    \item A key recommendation in this theme is to use tools that target or support the DevSecOps paradigm (e.g., hybrid tools [S5], continuous vulnerability assessment tools [S6]). 
    \item Studies also recommend moving to cloud solutions if it reduces the drawbacks of in-house tool usage, which affect DevSecOps goals (e.g., SAST) [S4].
    \item A range of methods and guidelines were reported to address the vulnerabilities and security limitations of tools and the CDP in DevSecOps [S6-S10].
\end{itemize}

\end{tcolorbox}

\subsubsection{Solutions proposed related to Practices}
This section reports the solutions related to practices.

\paragraph{S11: Adapting standards, policies, models, service level agreements into testable criteria}
A frequently reported challenge was the inability to automate certain manual security practices to fit into a DevSecOps pipeline [Ch10]. To overcome this problem, authors attempted to fully or partially automate practices such as compliance testing, risk management, security, and privacy by design [P02, P39, P47]. One of the key proposals reported in these studies was adapting standards, policies, models, and service level agreements (SLA) into testable criteria. For example, SLAs have been used to devise a security-by-design methodology that can be integrated with modern paradigms such as DevOps [P02]. In this case [P02], the SLAs were used to model and assess security requirements. SLAs were also reported to be used for security and privacy assurance in multi-cloud systems [P37, P39]. Here, the focus was to make SLAs machine-readable and based on security and privacy standards [P39].

To perform compliance testing practices at DevOps speeds, developers can use industry tools to adapt information security standards into testable criteria [P08]. Studies have also reported work on continuous automated compliance testing mechanisms, which would suit the DevSecOps paradigm [P47].  

\paragraph{S12: Automated vulnerability detection through requirement analysis}
Studies have proposed methods to automate security requirement analysis to suit CSE paradigms. For example, toolchains have been developed to automatically translate requirements documents into ontological models, analyze those models, and report the results [P35]. Once corrective guidance can be obtained from toolchains, developers are able to address security issues very early in the process.

\paragraph{S13: Devising security metrics or metric-based approaches}

In several empirical studies, developers discussed metrics for software security assessment used in their organizations. One proposal regarding metrics was to measure second-order effects in the development process [P20]. The following are some of the metrics discussed based on [P04] and [P20].

\begin{itemize}
    \item The number of developers modifying the file or with security training 
    \item The number of commits per time period
    \item The number of mistakes uncovered from known security classifications (e.g., Open Web Application Security Project (OWASP) top 10)
    \item The time spent to correct mistakes, category-wise (if a known classification was used)
    \item Internal/external penetration testing results for systems.
\end{itemize}
 
A taxonomy of metrics was developed by Diaz et al. [P38] relating to monitoring IoT environments.

\paragraph{S14: Effective process documentation and logging strategies}
Due to the rapid pace of software releases in the DevOps paradigm, practitioners do not prioritize documentation and logging processes. However, these are some of the best practice processes in software development and must be followed irrespective of release speed [P06]. For example, if the automated deployment and testing outcomes are documented, these will serve as evidence for audits [P06]. Therefore, studies have discussed using suitable tools and maintaining the required data (e.g., security logging of user access, metadata, and document repositories) to conduct documentation processes in a rapid manner to suit the DevOps paradigm [P06]. 

\paragraph{S15: Big data and behavioral analytics techniques}
As a solution to the reported limitations of the traditional data gathering methods in continuous practices, new approaches have been presented. For example, behavioral analytics techniques can be combined with big data solutions to obtain the advantages of these new technologies in DevSecOps [P15]. The main objective is to obtain fast feedback from the end-user of systems that are rapidly updated using DevSecOps practices. The following specific advantages can also be achieved [P15].

\begin{itemize}
\item Using user profiling to be aware of actual user behavior where customized responses can be provided from the application 
\item Predictive analytics to be aware of trends in user behaviors and intentions 
\item Setting up benchmarks to capture various measurements
\end{itemize}

\paragraph{S16: Shifting security to the left}
Shifting security activities and practices to the left of the development cycle is a key recommendation in DevSecOps [P04]. This shift would assign a higher priority to security practices at a very early stage of the development process [P08]. As a result, developers are able to identify security issues such as vulnerabilities early. This would overcome costly security fixes at the later stages of the cycle [P04].

This solution contains a combination of recommendations reported in other themes (i.e., People and Tools). Firstly, to conduct security activities at the initial stages of the cycle, security team members need to be involved at those stages [P08]. Secondly, the correct tool-sets need to be in place early [P16]. 

\paragraph{S17: Implementing continuous security assessment practices}
Continuous security assessment is another critical recommendation in DevSecOps. In this practice, security is treated as a key concern across all stages of the development process and even in the post-deployment period [P14].  It is a smart and lightweight approach to identifying security vulnerabilities in a rapid deployment environment [P14]. Similar to shifting security to the left, continuous security practices require security team members to be involved in the early stages of the cycle and continue to do so in a continuous manner. For this to occur, there needs to be a consensus on how security practices would be included in the process [P16]. Next, the required tools need to be in place [P12]. 

As an example of continuous security assessment practices, Continuous Monitoring (CM) can be noted [P07, P38]. CM is a recommended practice in highly regulated environments [P07]. In such conditions, all environmental settings, events, logs, alerts, and assets should be continuously monitored. CM was also recommended in IoT systems to obtain fast and continuous feedback from the operations to development teams [P38].

\paragraph{S18: Security patch management using DevOps practices}
Another aspect of DevOps and security is the usage of DevOps practices to address security vulnerabilities rapidly [P20]. Due to rapid releases in DevOps, vulnerabilities might exist in production code. Therefore, once vulnerabilities are detected, it is important to address them (e.g., security patches) as soon as possible. For this purpose, the continuous practices of DevOps (i.e., CD/CDE) can be used to deliver security patches rapidly.

\paragraph{S19: Using threat analysis practices}
Threat modeling is a practice used to identify, communicate and understand threats and mitigation methods \cite{owasp}. Moreover, it can be used to identify vulnerabilities at various stages and outputs. For example, to detect vulnerabilities in a CD pipeline, the STRIDE threat modeling approach can be used [P50].

\subsubsection{Summary of the solutions related to practices}

\begin{tcolorbox}[left=1pt, top=1pt, right=1pt, bottom=1pt]
\begin{itemize}
    \item In DevSecOps, security should be treated as a key concern from the start of the process (\textit{shifting left}), and it should continue to be so, throughout the cycle (\textit{continuous security}) [S16, S17]. 
    \item  We have reported on how automation of certain practices (e.g., SLAs made machine-readable), security metrics and other strategies could be used to achieve the above noted goals [S11-S19].

\end{itemize}

\end{tcolorbox}

\subsubsection{Solutions proposed related to Infrastructure}
This section presents the reported solutions based on various challenging infrastructures.  

\paragraph{S20: Strict access management and policies}
A key recommendation of DevSecOps is to enable multiple team members of different teams (Dev/Sec/Ops) to work on the same pipeline. In practice, it entails giving permission and various access rights to these members. However, strict access management policies are proposed if the output is released to a sensitive and highly regulated environment. For example, need-based access (based on the principle of least privilege) is recommended for developers who need access to highly regulated environments or settings [P07]. Also, changes to the production environment need to be automated, which should result in removing production access from developers [P07]. 

\paragraph{S21: Adopting Infrastructure as code}
In complex and highly regulated infrastructures, IaC is highly recommended. As IaC enables infrastructure to be versioned, tested, built, and deployed, this is seen as a suitable solution to manage complex environments with security concerns (e.g., air-gapped secure environments [P07]). IaC also supports environment parity by setting up pre-configured systems and networks (as relating to highly regulated environments [P27]). Stakeholders should agree to these configuration settings of the environment at the inception of a project. When a need arises, they could then use IaC to build and deploy an environment. This addresses the challenge of the lack of centralized processes to deploy software and environmental settings in highly regulated environments [P27]. However, in using IaC tools, developers need to be mindful of the security limitations of the tool itself [Ch8].

\paragraph{S22: Creating simulation or replication environments for testing}
Setting up simulation environments is a recommended solution in complex infrastructures for the purpose of testing [P27]. By setting up common simulation environments, internal testing can be carried out by developers in these infrastructures. This could be very useful in terms of conducting security-related testing and gathering end-user feedback. The ENACT framework for trustworthy IoT systems by Ferry et al. [P19] also contains a test and simulation tool in one of its layers. These tools enable testing application scenarios against programmed circumstances [P19].

\paragraph{S23: Model-driven engineering to support DevSecOps}
A model-driven approach was reported to address challenges related to IoT infrastructures [P10]. The \textit{model once, generate anywhere} quality of this approach enables to tackle specific challenges of an IoT infrastructure, such as high heterogeneity. Further, this approach can address the security and privacy difficulties in IoT. It provides the methods to specify security and privacy requirements and supports the automatic deployment and the relevant mechanisms (e.g., GENESIS, an approach that leverages model-driven engineering to support the DevSecOps approach in IoT infrastructures [P10]).

\paragraph{S24: Systematic evaluation of product-specific vulnerabilities}
This is a solution proposal targeting highly regulated environments (e.g., medical devices [P13]). Due to heavy regulations, the vulnerability handling activity needs to be highly systematic and transparent [P13]. 
Further, design aware risk assessment is critical for these industries (e.g., design of the system in a highly regulated setting). In today's context, software consists of a number of diverse components (e.g., open-source software). Therefore, in critical environments, the vulnerabilities of the components used need to be assessed systematically.

\paragraph{S25: Hybrid life cycles with data-security focus}
With the cloud environment getting more complex and novel (e.g., multi-cloud setups), ensuring data security has become challenging while rapidly deploying software. To address this issue, a particular focus needs to be paid to data security in parallel to the development cycle [P18]. A joint Software as a Service (SaaS) security life cycle was proposed, which combined the data security and software development life cycles to address this specific issue [P18].

\paragraph{S26: Framework support for DevSecOps}
Several reviewed studies have proposed specific frameworks to address the security challenges in various challenging infrastructures.

A systems-of-systems security framework was proposed targeting the requirements definition phase for cloud systems [P03]. This research is an example of a case study for the implementation of DevSecOps in SoS [P03]. Framework support for multi-cloud application modeling was reported by two in-scope studies [P37], [P39]. Details of the MUSA security modeling language and supporting tool for multi-cloud applications was reported by [P37]. This paper introduced the \textit{MUSA SecDevOps framework} which included the above components. We reported the machine-readable SLA mechanism of this project (MUSA) in the solutions related to the practices section [P39] [S11]. A secure DevOps framework was proposed, which used Network Functions Virtualization (NFV) and micro-service pattern designs [P24]. This framework targeted distributed and heterogeneous cloud environments [P24]. The \textit{ENACT DevOps framework} targeted IoT systems [P19]. The goal of this framework was to establish trustworthy (i.e., preservation of security \& privacy) smart IoT systems. Trihinas et al. [P54] proposed the \textit{Unicorn Framework} to address the challenges of developing scalable and secure applications based on the microservices architecture in multi-cloud environments.

\subsubsection{Summary of the infrastructure-related solutions}

\begin{tcolorbox}[left=1pt, top=1pt, right=1pt, bottom=1pt]
\begin{itemize}
    \item This theme captured several framework based solutions for the infrastructure challenges reported. The frameworks provide holistic solutions targeting DevSecOps adoption in various challenging infrastructures [S26]. 
    \item In addition to the frameworks, strict access management [S20], model-driven engineering [S23] and setting up simulation environments [S22] were some of the other solutions specific for the infrastructure theme synthesized.
\end{itemize}

\end{tcolorbox}

\subsubsection{Solutions to people-centric challenges}
This section reports the solutions based on the \textit{People} theme.

\paragraph{S27: Facilitating inter-team communication and collaboration with the appropriate controls or standards}
Several studies reported inter-team collaboration (as opposed to working in silos) as a best practice that should be adopted in DevSecOps [P06, P42]. One related recommendation is the formation of multidisciplinary teams in an organization [P08]. This would lead to a high level of collaboration among development, security, and operations team members on a continual basis starting from the early stages of the process. Other recommendations include customizing security tools so that the feedback cycle with other teams is short [P42] and increasing developer engagement in security tasks (e.g., security incident management) [P41]. 

Despite the above encouragement for more collaboration, this should be done with the appropriate controls or standards. For example, a clear separation of duties should be established for various cross-team members [P06]. Also, there should be a clear established way of communication between team members. Manual communication methods such as emails should be discouraged as it would add up to the communication effort [P06]. Automated methods should instead be utilized to inform team members of activities occurring in the processes. For instance, the relevant stakeholders should be automatically notified from a system about testing efforts, successful installations, etc., rather than an administrator manually messaging each person [P06].

Studies have also reported the need for a systematic, consistent and transparent methodology for the communication processes, especially in regulated or restricted environments. For example, the communication of risk information and evaluation results or rationales to stakeholders on a continuous basis is required in such environments [P13]. For this purpose, verified key stakeholders and communication channels need to be in place [P27].

\paragraph{S28: Having security champions in teams}
In a DevSecOps environment, there are different teams with differing priorities. Therefore, the bridging role of security champions is recommended for this environment [P04]. Security champions are security-minded developers who typically have the most security training in a team [P04]. Therefore, they treat security as a priority. Due to this reason, they can act as a bridge between development and security teams. Another benefit of this role is the effect of security champions on developers' resistance to security activities. As these members are usually from the same development team, programmers are less likely to perceive security champions as outside agents hindering progress [P20].

\paragraph{S29: Carrying out organizational Human Resource Management (HRM) programs in parallel} 
The culture change for DevSecOps can be challenging for certain people in an organization. To tackle this issue, studies recommended carrying out HRM programs in parallel to DevSecOps transformation projects [P03]. These programs should target common sentiments such as fear of being replaced or being recognized and losing control of one's own work [P03].

\paragraph{S30: Implementing security knowledge sharing methods and training}

Implementation of security training and knowledge sharing methods could play an important role in improving the security awareness of team members to carry out relevant tasks. For example, security-related knowledge is vital in setting up and using suitable security tools. If the development team is using static analysis tools to check for vulnerabilities, they would need the relevant knowledge to identify actual security issues and false positives [P04]. Further, as it is practically difficult for every developer to be a security expert, it is important that they are able to recognize when they would benefit from the advice of such an expert [P20]. Hence, at least some basic level of security knowledge or awareness is needed for this purpose.

Studies reported detail on specific security training activities such as completing online coursework, participating in developer boot camps and in-house security awareness sessions [P42]. \textit{Blameless security retrospectives} are another type of activity highlighted [P04]. What is advocated in this process is that if security issues are discovered, they are not seen as faults of a certain person. Therefore, the focus is to identify issues and share knowledge.

\paragraph{S31: Integrity protection frameworks}
To address insider threats, the authors proposed system and data integrity protection frameworks. For example, Ahmadvand et al. [P29] formed a set of integrity protection requirements based on the threats and then proposed an integrity protection framework that targets holistic integrity protection in microservice-based systems.

\subsubsection{Summary of the solutions to people-centric challenges}

\begin{tcolorbox}[left=1pt, top=1pt, right=1pt, bottom=1pt]
\begin{itemize}
    \item The role of security champions is a key recommendation for DevSecOps proposed by the reviewed studies [S28]. This role facilitates other advice of the People theme, such as continuous and frequent inter-team collaboration [S27]. 
    \item However, studies highlight the need for controlled and standardized communication strategies for DevSecOps teams [S27]. 
    \item Finally, security knowledge sharing and training activities are crucial in DevSecOps as developers are required to carry out security tasks [S30].
\end{itemize}

\end{tcolorbox}

\subsection{DevSecOps support in practice}

In this section, we describe an example of how our results can be combined to offer DevSecOps support in practice. The concrete example that we provide is \textit{support for setting up a DevOps workflow with suitable security controls (e.g., application security tools) integration}.

One of the main challenges in DevSecOps practice is integrating security controls (i.e., tools and practices) into the DevOps workflow without negatively impacting the rapid delivery goals [Ch14]. Further, while there are many security assessment and control methods, in an environment with rapid deployment needs, adopting all these solutions can overwhelm developers [Ch1]. Therefore, proper support for DevSecOps in practice would guide developers on selecting and applying the most suitable security controls while having a minimal effect on the deployment frequencies in DevOps. This requires an appropriate balance between speed and security needs to be achieved for successful DevSecOps adoption. We offer some practical suggestions on how this balance can be achieved in practice.

Our results show that development teams should set up a suitable DevOps workflow from the inception of the project. For this purpose, developers should be encouraged to integrate security tools that facilitate the rapid deployment of software into this workflow. We propose the usage of hybrid tools (i.e., IAST: Interactive Application Security Testing) that combine the advantages of traditional application security testing tools (e.g., [S5]). However, due to other practical reasons (e.g., cost, developer preference), traditional tools may also be required to be used in the DevOps workflow. In such cases, practitioners should plan to minimize the negative drawbacks of such tools using some of our proposed solutions. For example, developers can set up a parallel testing pipeline in the workflow for static scanning using a time-consuming SAST tool [S3]. In this setup, the main pipeline moves forward while deep scans occur in parallel. This would ease some of the time constraints in a rapid deployment environment.

Another method of obtaining DevSecOps support is to use an application security testing as a service provider for specific testing activities (e.g., static analysis as a service [S4]).  The reduced time to obtain the results of this model is suited for DevOps. Further, if the required in-house security expertise is not available, obtaining such services would increase the speed of conducting the security testing activity with higher accuracy levels. 

Therefore, by adopting a combination of the above-noted solutions, practitioners would be able to better balance the speed of deployments in a DevOps workflow while ensuring the security of the outputs.

\subsection{RQ3: What are opportunities for future research or gap areas for technological development (e.g., tool support) or framework support in this domain?}

\begin{landscape}

\begin{figure}[t]
    \includegraphics[scale=.75]{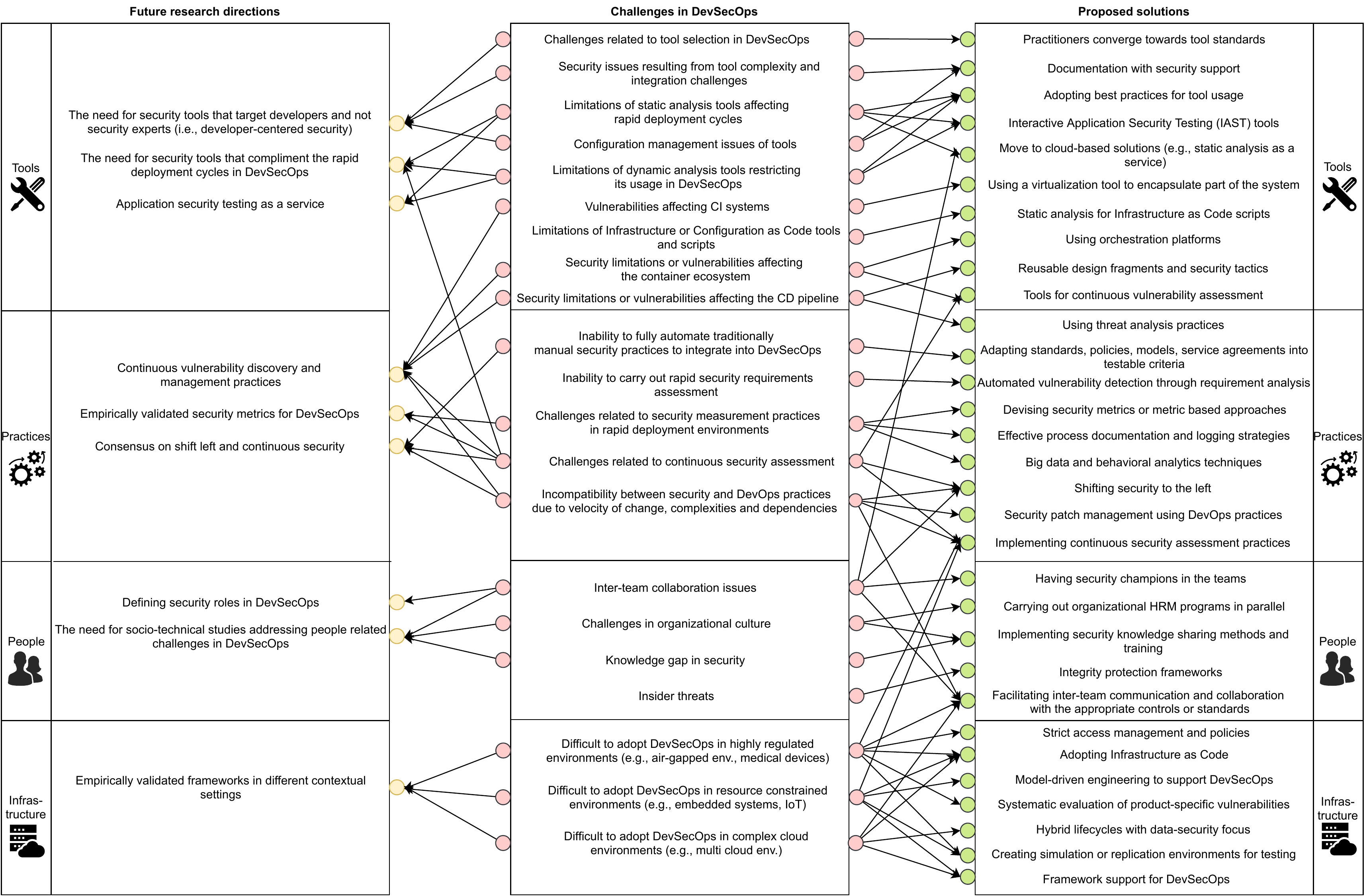}
    \caption{Mapping of challenges, solutions and gap areas devised in the study}
    \label{fig:map}
\end{figure}

\end{landscape}

In this section, we discuss the gap areas for future research in this domain using the results of RQ1 and RQ2. We provide the mapping of our study results (Figure \ref{fig:map}) to enable the reader to quickly determine the association between challenges, proposed solutions, and gap areas in a holistic manner. 

\subsubsection{The need for security tools that complement the rapid deployment cycles in DevSecOps}

SAST and DAST tools are popular technologies in the industry, which are used in various stages in the development process. However, several studies reported challenges related to integrating these established tools in the rapid deployment cycles [Ch4, Ch5]. While IAST was mentioned as a new type of hybrid tool to address some of these challenges [S5], very limited research exists on these technologies. Therefore, hybrid tools that combine the features of existing technologies in a way that is suited for modern paradigms such as DevSecOps are a clear gap area for further research.

Another key recommendation proposed in our study is using tools that make continuous security assessment possible [S6]. However, there were only a few studies that reported empirically validated solutions for such tools. This is also a gap area for new tool development. 

\begin{tcolorbox}[left=1pt, top=1pt, right=1pt, bottom=1pt]
\textbf{Gap area:} We highlight the need for application security testing tools and methods, which specifically cater to the nature of the DevSecOps paradigm (e.g., rapid CI/CD cycles). 
\end{tcolorbox}

\subsubsection{Application security testing as a service}

Another solution proposal that is compatible with DevSecOps is the usage of application security testing tools as a service. Many industry resources have advocated for this approach, with cloud-native applications on the rise (e.g., \cite{lemos}). Also, many leading companies in the application security testing market are now providing their products as a service. However, we could only capture one peer-reviewed study that proposed application security testing as a service in our SLR [P25]. This study demonstrated how static analysis offered as a service could overcome the challenges that restrict this tool's usage in DevOps [S4]. 
\begin{tcolorbox}[left=1pt, top=1pt, right=1pt, bottom=1pt]
\textbf{Gap area:} Further research can be carried out on how other application security testing methods (e.g., DAST) can be offered as a service to suit the DevSecOps paradigm.
\end{tcolorbox}

\subsubsection{The need for developer-centered security solutions}

Developers, similar to end-users in security tasks, need support to create applications that are secure \cite{tahaei2019survey}. One of the main difficulties that developers face is the use of complicated security tools in their development workflow, resulting in low adoption of such tools \cite{tahaei2019survey}. 

We reported several tool-related challenges on selection, configuration, and integration that affected DevSecOps [Ch1, Ch2, Ch3]. The limitations in the documentation for these tools made the challenge more critical [Ch2]. The root cause for such problems is that most of the traditional security tools are developed for the use of security experts and their workflow. Therefore, developers are finding it challenging to select such tools from a large range of existing products [Ch1] and then use them in their own workflow [Ch2, Ch3]. This is an obstacle in modern paradigms such as DevSecOps, as the developer must carry out some security-related tasks. 

\begin{tcolorbox}[left=1pt, top=1pt, right=1pt, bottom=1pt]
\textbf{Gap area:} We advocate for empirically evaluated tools which target developers in their development workflow (i.e., developer-centered security solutions). 
\end{tcolorbox}

\subsubsection{Continuous vulnerability discovery and management practices}
Our results presented many challenges related to vulnerabilities of the tools and pipeline [Ch6, Ch7, Ch9]. However, continuous vulnerability assessment processes were not widely adopted in the industry [Ch13]. Therefore, more research is needed to discover the reasons for this low adoption. 

After the vulnerabilities are discovered, the next challenge is the efficient management of the vulnerabilities within the rapid deployment cycles. We reported several challenges regarding this practice too [Ch14]. Due to the substantial effort needed from engineers to address and manage the vulnerabilities (e.g., determine true/false positives, rectify true positives), they face a trade-off between speed and security. 

\begin{tcolorbox}[left=1pt, top=1pt, right=1pt, bottom=1pt]
\textbf{Gap area:} Developers need to be presented with specific guidance on how the trade-off between DevOps release speeds and security of the software can be effectively balanced in a practical setting.
\end{tcolorbox}

\subsubsection{Consensus on shift left and continuous security}

Shift left security and continuous security assessment were two key solutions related to practices reported in our SLR. There is a high interest among DevOps practitioners to implement these practices in their organizations \cite{Gitlab2020}. However, a consensus is required on how these practices can be practically implemented and validated (e.g., what are the processes to be shifted or automated and to which point in the pipeline?). 

For example, studies reported that security tasks should be moved to the left and automated as much as possible [S16]. However, more work needs to be done on how the traditionally manual security practices can be automated to fit into DevSecOps. Concerning this particular challenge, there were only a few solutions captured by our SLR [S11], even though many such security practices were reported as challenges in the literature [Ch10]. For example, practices such as threat modeling, penetration testing, and code review were not addressed.

\begin{tcolorbox}[left=1pt, top=1pt, right=1pt, bottom=1pt]
\textbf{Gap area:} More work is needed to establish consensus on proposals such as shift left security and how traditionally manual security practices can be automated to suit the DevSecOps paradigm.
\end{tcolorbox}

\subsubsection{Empirically validated security metrics for DevSecOps}

Studies note that measuring security in software is a very difficult task [P20]. This task is even more challenging in DevSecOps, where multiple cross-disciplinary teams aim to deliver software rapidly. To address this problem, models such as Building Security In Maturity Model (BSIMM) have been devised in the industry {\cite{Migues2020}}. Organizations use BSIMM as a measurement tool to compare their software security initiatives with the data from the broader BSIMM community. Such models emphasize the importance of security metrics for software security assessment. For example, in BSIMM, the importance of using metrics for measurement is highlighted (e.g., for data-driven governance and track performance). The BSIMM report also states that there is a high interest in the industry to consume real-time security events to produce useful metrics {\cite{Migues2020}}. However, our results indicate a lack of empirically validated security metrics in this domain [Ch12]. Therefore, this is a key area for future work due to the lack of research and potential importance to DevSecOps.

\begin{tcolorbox}[left=1pt, top=1pt, right=1pt, bottom=1pt]
\textbf{Gap area:}  While researchers have stated the importance of introducing security metrics into DevSecOps practices, a very limited number of studies have established such metrics by empirically validating them.  
\end{tcolorbox}

\subsubsection{Defining security roles in DevSecOps}

In DevSecOps, the traditional roles of developers and security engineers have changed. With the shift left approach promoted in DevSecOps, the developers are encouraged to carry out security-related tasks. The security engineer's role has also changed from being involved in a specific stage to all stages of the cycle. However, ambiguity lies about \textit{who makes which security decisions} in a practical scenario. Further to the above, several studies have stated the importance of bridging roles such as \textit{security champions} for DevSecOps [S28]. However, the exact role of such developers with regard to security decision-making has not been clearly established.

\begin{tcolorbox}[left=1pt, top=1pt, right=1pt, bottom=1pt]
\textbf{Gap area:}  More effort needs to be directed at clearly defining the security roles of the team members in DevSecOps.
\end{tcolorbox}

\subsubsection{The need for socio-technical studies addressing people-related challenges in DevSecOps}

The paradigms of DevOps and DevSecOps are widely recognized as a cultural transformation for an organization \cite{sanchez2018characterizing}, \cite{Atlassian.com}. However, in our study, the \text{people} theme captured the least amount of studies in our search. We also noticed that the challenges reported in the people theme had a wide-ranging effect across themes. For example, developers lacking security skills resulted in challenges in the tools and practices themes (e.g., Ch04, Ch13).  

\begin{tcolorbox}[left=1pt, top=1pt, right=1pt, bottom=1pt]
\textbf{Gap area:} We highlight the need for socio-technical studies which focus on the people-centric challenges and their effect on the other components of DevSecOps. (e.g., tool usage, implementation of practices).
\end{tcolorbox}

\subsubsection{Empirically validated frameworks in different contextual settings}
The infrastructure theme captured several framework-based solutions for the challenges reported. Most of these frameworks were presented as case studies or were not adequately evaluated [S26]. Some authors presented the frameworks as solution proposals for their own context (e.g., company) [P03]. Therefore, more empirical studies are needed for the broader applicability of these solutions and associated components (e.g., modeling languages), especially in different contextual settings.

\begin{tcolorbox}[left=1pt, top=1pt, right=1pt, bottom=1pt]
\textbf{Gap area:} Empirically validated solutions (e.g., frameworks) which enable adopting DevSecOps in complex, resource constrained, or highly regulated environments is a current need in this domain.
\end{tcolorbox}

\section{Threats to validity}

Similar to previous systematic reviews \cite{unterkalmsteiner2011evaluation, khatibsyarbini2018test} our results may be affected by the potential threats associated with imperfect collection of primary studies, bias in study selection/extraction, and publication bias.

\paragraph{Missing primary studies}
We are unable to guarantee that we have captured all the relevant primary studies in our SLR. This is due to limitations in the search method, \cite{dybaa2008empirical}, and non-comprehensive venues or databases \cite{zhou2016map}. We aimed to minimize this effect as follows.

Firstly, we ran pilot searches on frequently used index engines and publisher sites, evaluated the results, and selected the sources for our papers. The search string was then iteratively improved to capture the relevant studies. In determining the sources, we chose one index engine and two individual publishers for our study. While this approach results in duplicates, it reduces the threat of missed studies \cite{unterkalmsteiner2011evaluation}. We then performed backward, forward snowballing \cite{wohlin2014guidelines} and snowballed the primary study lists of other secondary studies (Table \ref{table:other_reviews}) to capture the missed papers. From the above activities, we attempted to minimize the number of relevant studies missed by our search. 

\paragraph{Bias in study selection and data extraction} Bias in study selection (i.e., applying the inclusion/exclusion criteria) and data extraction are common limitations that affect SLRs \cite{zhou2016map}. To reduce this effect, we developed a study protocol that defined the research questions and search procedures, inclusion/exclusion criteria, and a data extraction strategy. \cite{dybaa2008empirical, unterkalmsteiner2011evaluation}. A well-defined protocol is said to increase the consistency in the selection of primary studies and data extraction \cite{unterkalmsteiner2011evaluation}. This document was discussed and shared among all authors. Two authors were engaged in the data extraction process using data extraction sheets. These sheets were stored in shared folders and checked by the other authors. Regular discussions were held among the authors to assess the outcome of these stages and to cross-check the completeness of the search and data extraction \cite{zhou2016map}.

\paragraph{Publication bias} This refers to the issue that positive results are more likely to be reported compared to negative results \cite{Kitchenham07guidelinesfor}. However, in our study, we captured many papers which reported negative effects, such as challenges faced by practitioners in this domain (RQ1). When forming the gaps (RQ3), these challenges were assessed against the solutions (RQ2). Therefore, unreported negative results would only have a moderate effect on these contributions.

\section{Conclusion}
Based on our results, we conclude:

\begin{itemize}
    \item Out of the themes devised in this study, the main focus area for researchers in DevSecOps is automation and tool usage. We found that some of the older technologies, such as SAST and DAST tools had drawbacks that affected DevSecOps goals. Therefore, research and development on new technologies that support the rapid deployment cycles of DevOps (e.g., hybrid tools or IAST) is a current need.
    \item A large number of tool-related security issues and vulnerabilities were captured by our study (e.g., vulnerabilities affecting containers). Therefore, developers should take appropriate action to minimize the effect of these security issues when including such tools in the pipeline. 
    \item Shift-left security and continuous security assessment were two key recommendations related to the practices-theme of our study. These practices keep security as a priority in a continuous manner throughout the deployment cycle. However, this is another area where tool support is lacking (e.g., tools that support continuous security assessment).
    \item Inability to automate traditionally manual security practices is a critical problem in this field. These practices are hard to be fully integrated with the CD or CDE practices of DevOps. Further studies are needed on how a suitable trade-off can be achieved in balancing the goals of DevOps and such security practices.
    \item Even though cultural or human aspects are critical for DevSecOps success, these are less discussed areas in the literature. In our study, the people-related challenges had a wide-ranging effect across the other themes. This leads us to recommend more socio-technical related research exploring the people-related challenges and solutions in this domain.
    \item Adopting DevSecOps principles or practices in various complex, resource-constrained, and highly regulated infrastructures is a growing area of research. However, more empirically evaluated solutions are needed to ensure wider adoption of such tools or frameworks.
    \item Finally, we note that many DevSecOps adoption challenges and solutions captured in our study were interrelated across themes. We provided the mapping of the results of our study to illustrate this point. Therefore, researchers need to thoroughly consider challenges or solutions across themes and their applicability when devising studies in this domain.
\end{itemize}

\section{Acknowledgement}
The work has been supported by the Cyber Security Research Centre Limited whose activities are partially funded by the Australian Government’s Cooperative Research Centres Programme. We also thank our colleagues at CREST and the anonymous reviewers for their useful feedback on this work.

\vspace{-1cm}

\renewcommand{\refname}{\section{References}}

\bibliographystyle{elsarticle-num}

\bibliography{mybibfile.bib}

\begin{thebibliography}{10}
\expandafter\ifx\csname url\endcsname\relax
  \def\url#1{\texttt{#1}}\fi
\expandafter\ifx\csname urlprefix\endcsname\relax\def\urlprefix{URL }\fi
\expandafter\ifx\csname href\endcsname\relax
  \def\href#1#2{#2} \def\path#1{#1}\fi

\bibitem{leite2019survey}
L.~Leite, C.~Rocha, F.~Kon, D.~Milojicic, P.~Meirelles, A survey of devops
  concepts and challenges, ACM Computing Surveys (CSUR) 52~(6) (2019) 1--35.

\bibitem{bass2015devops}
L.~Bass, I.~Weber, L.~Zhu, DevOps: A software architect's perspective,
  Addison-Wesley Professional, 2015.

\bibitem{mann2018state}
A.~Mann, A.~Brown, M.~Stahnke, N.~Kersten, State of devops report 2018, Tech.
  rep. (2018).

\bibitem{SignalSciences2020}
{Signal Sciences}, {A DevOps roadmap for Security: Third Edition}, Tech. rep.
  (2020).

\bibitem{riungu2016devops}
L.~Riungu-Kalliosaari, S.~M{\"a}kinen, L.E. Lwakatare, J.~Tiihonen,
  T.~M{\"a}nnist{\"o}, Devops adoption benefits and challenges in practice: a
  case study, in: International Conference on Product-Focused Software Process
  Improvement, Springer, 2016, pp. 590--597.

\bibitem{myrbakken2017devsecops}
H.~Myrbakken, R.~Colomo-Palacios, Devsecops: a multivocal literature review,
  in: International Conference on Software Process Improvement and Capability
  Determination, Springer, 2017, pp. 17--29.

\bibitem{flechais2005designing}
I.~Fl{\'e}chais, Designing secure and usable systems, PhD diss., University
  College London.

\bibitem{Shiftleft2021}
\href{https://www.techarcis.com/shift-left-testing-explained-by-sunil-sehgal-of-techarcis-2/}{What
  is shift left testing?} (2021).
\newline\urlprefix\url{https://www.techarcis.com/shift-left-testing-explained-by-sunil-sehgal-of-techarcis-2/}

\bibitem{Sharma2020}
R.~Sharma,
  \href{https://www.netsolutions.com/insights/what-is-devsecops/}{{What is
  DevSecOps? Definition, Importance, Benefits, Challenges, and Best Practices}}
  (2020).
\newline\urlprefix\url{https://www.netsolutions.com/insights/what-is-devsecops/}

\bibitem{howard2006process}
M.A. Howard, A process for performing security code reviews, IEEE Security \&
  privacy 4~(4) (2006) 74--79.

\bibitem{peterson2020}
J.~Peterson,
  \href{https://resources.whitesourcesoftware.com/blog-whitesource/dast-dynamic-application-security-testing}{Dynamic
  application security testing: Dast basics} (2020).
\newline\urlprefix\url{https://resources.whitesourcesoftware.com/blog-whitesource/dast-dynamic-application-security-testing}

\bibitem{Ng2018}
A.~Ng,
  \href{https://www.cnet.com/news/equifaxs-hack-one-year-later-a-look-back-at-how-it-happened-and-whats-changed/}{{How
  the Equifax hack happened, and what still needs to be done}} (2018).
\newline\urlprefix\url{https://www.cnet.com/news/equifaxs-hack-one-year-later-a-look-back-at-how-it-happened-and-whats-changed/}

\bibitem{Bushwick2020}
S.~Bushwick,
  \href{https://www.scientificamerican.com/article/giant-u-s-computer-security-breach-exploited-very-common-software/}{{Giant
  U.S. Computer Security Breach Exploited Very Common Software}} (2020).
\newline\urlprefix\url{https://www.scientificamerican.com/article/giant-u-s-computer-security-breach-exploited-very-common-software/}

\bibitem{mann2019state}
A.~Mann, A.~Brown, M.~Stahnke, N.~Kersten, State of devops report 2019, Tech.
  rep. (2019).

\bibitem{Prince2016}
S.~Prince,
  \href{https://www.mindtheproduct.com/what-the-hell-are-ci-cd-and-devops-a-cheatsheet-for-the-rest-of-us/}{{The
  Product Managers' Guide to Continuous Delivery and DevOps}} (2016).
\newline\urlprefix\url{https://www.mindtheproduct.com/what-the-hell-are-ci-cd-and-devops-a-cheatsheet-for-the-rest-of-us/}

\bibitem{shahin2017continuous}
M.~Shahin, M.A. Babar, L.~Zhu, Continuous integration, delivery and deployment:
  a systematic review on approaches, tools, challenges and practices, IEEE
  Access 5 (2017) 3909--3943.

\bibitem{shahin2019empirical}
M.~Shahin, M.~Zahedi, M.A. Babar, L.~Zhu, An empirical study of architecting
  for continuous delivery and deployment, Empirical Software Engineering 24~(3)
  (2019) 1061--1108.

\bibitem{checkmarx_2020}
Checkmarx,
  \href{https://www.checkmarx.com/ebooks/an-integrated-approach-to-embedding-security-into-devops}{An
  Integrated Approach to Embedding Security into DevOps}, 2020.
\newline\urlprefix\url{https://www.checkmarx.com/ebooks/an-integrated-approach-to-embedding-security-into-devops}

\bibitem{fitzgerald2017continuous}
B.~Fitzgerald, K.J. Stol, Continuous software engineering: A roadmap and
  agenda, Journal of Systems and Software 123 (2017) 176--189.

\bibitem{bosch2014continuous}
J.~Bosch, Continuous software engineering: An introduction, in: Continuous
  software engineering, Springer, 2014, pp. 3--13.

\bibitem{schermann2016empirical}
G.~Schermann, J.~Cito, P.~Leitner, U.~Zdun, H.~Gall, An empirical study on
  principles and practices of continuous delivery and deployment, Tech. rep.,
  PeerJ Preprints (2016).

\bibitem{zahedi2020mining}
M.~Zahedi, R.N. Rajapakse, M.A. Babar, Mining questions asked about continuous
  software engineering: A case study of stack overflow, in: Proceedings of the
  Evaluation and Assessment in Software Engineering, 2020, pp. 41--50.

\bibitem{stahl2017continuous}
D.~St{\aa}hl, T.~M{\aa}rtensson, J.~Bosch, Continuous practices and devops:
  beyond the buzz, what does it all mean?, in: 2017 43rd Euromicro Conference
  on Software Engineering and Advanced Applications (SEAA), IEEE, 2017, pp.
  440--448.

\bibitem{leppanen2015highways}
M.~Lepp{\"a}nen, S.~M{\"a}kinen, M.~Pagels, V.P. Eloranta, J.~Itkonen, M.V.
  M{\"a}ntyl{\"a}, T.~M{\"a}nnist{\"o}, The highways and country roads to
  continuous deployment, Ieee software 32~(2) (2015) 64--72.

\bibitem{chen2015continuous}
L.~Chen, Continuous delivery: Huge benefits, but challenges too, IEEE Software
  32~(2) (2015) 50--54.

\bibitem{mohan2016secdevops}
V.~Mohan, L.~ben Othmane, Secdevops: Is it a marketing buzzword?-mapping
  research on security in devops, in: 2016 11th international conference on
  availability, reliability and security (ARES), IEEE, 2016, pp. 542--547.

\bibitem{bird2016devopssec}
J.~Bird, DevOpsSec: Securing software through continuous delivery, O'Reilly
  Media, 2016.

\bibitem{prates2019devsecops}
L.~Prates, J.~Faustino, M.~Silva, R.~Pereira, Devsecops metrics, in:
  EuroSymposium on Systems Analysis and Design, Springer, 2019, pp. 77--90.

\bibitem{sanchez2020security}
M.~S{\'a}nchez-Gord{\'o}n, R.~Colomo-Palacios, Security as culture: A
  systematic literature review of devsecops, in: Proceedings of the IEEE/ACM
  42nd International Conference on Software Engineering Workshops, 2020, pp.
  266--269.

\bibitem{mao2020preliminary}
R.~Mao, H.~Zhang, Q.~Dai, H.~Huang, G.~Rong, H.~Shen, L.~Chen, K.~Lu,
  Preliminary findings about devsecops from grey literature, in: 2020 IEEE 20th
  International Conference on Software Quality, Reliability and Security (QRS),
  IEEE, 2020, pp. 450--457.

\bibitem{rafi2020prioritization}
S.~Rafi, W.~Yu, M.A. Akbar, A.~Alsanad, A.~Gumaei, Prioritization based
  taxonomy of devops security challenges using promethee, IEEE Access 8 (2020)
  105426--105446.

\bibitem{rahman2015synthesizing}
A.A.U. Rahman, E.~Helms, L.~Williams, C.~Parnin, Synthesizing continuous
  deployment practices used in software development, in: 2015 Agile Conference,
  IEEE, 2015, pp. 1--10.

\bibitem{wettinger2015enabling}
J.~Wettinger, V.~Andrikopoulos, F.~Leymann, Enabling devops collaboration and
  continuous delivery using diverse application environments, in: OTM
  Confederated International Conferences" On the Move to Meaningful Internet
  Systems", Springer, 2015, pp. 348--358.

\bibitem{gotimer2016devops}
G.~Gotimer, T.~Stiehm, Devops advantages for testing: Increasing quality
  through continuous delivery, CrossTalk Magazine (2016) 13--18.

\bibitem{olszewska2015devops}
M.~Olszewska, M.~Wald{\'e}n, Devops meets formal modelling in high-criticality
  complex systems, in: Proceedings of the 1st international workshop on
  quality-aware DevOps, 2015, pp. 7--12.

\bibitem{wettinger2015dyn}
J.~Wettinger, U.~Breitenb{\"u}cher, F.~Leymann, Dyn tail-dynamically tailored
  deployment engines for cloud applications, in: 2015 IEEE 8th International
  Conference on Cloud Computing, IEEE, 2015, pp. 421--428.

\bibitem{shahin2017beyond}
M.~Shahin, M.A. Babar, M.~Zahedi, L.~Zhu, Beyond continuous delivery: an
  empirical investigation of continuous deployment challenges, in: 2017
  ACM/IEEE International Symposium on Empirical Software Engineering and
  Measurement (ESEM), IEEE, 2017, pp. 111--120.

\bibitem{kitchenham2004evidence}
B.A. Kitchenham, T.~Dyb{\aa}, M.~J{\o}rgensen, Evidence-based software
  engineering, in: Proceedings. 26th International Conference on Software
  Engineering, IEEE, 2004, pp. 273--281.

\bibitem{dyba2005evidence}
T.~Dyb{\aa}, B.A. Kitchenham, M.~J{\o}rgensen, Evidence-based software
  engineering for practitioners, IEEE software 22~(1) (2005) 58--65.

\bibitem{Kitchenham07guidelinesfor}
B.~Kitchenham, S.~Charters, Guidelines for performing systematic literature
  reviews in software engineering (2007).

\bibitem{laukkanen2017problems}
E.~Laukkanen, J.~Itkonen, C.~Lassenius, Problems, causes and solutions when
  adopting continuous delivery a systematic literature review, Information and
  Software Technology 82 (2017) 55--79.

\bibitem{chen2010towards}
L.~Chen, M.A. Babar, H.~Zhang, Towards an evidence-based understanding of
  electronic data sources, in: 14th International Conference on Evaluation and
  Assessment in Software Engineering (EASE), 2010, pp. 1--4.

\bibitem{dybaa2008empirical}
T.~Dyb{\aa}, T.~Dings{\o}yr, Empirical studies of agile software development: A
  systematic review, Information and software technology 50~(9-10) (2008)
  833--859.

\bibitem{wohlin2014guidelines}
C.~Wohlin, Guidelines for snowballing in systematic literature studies and a
  replication in software engineering, in: Proceedings of the 18th
  international conference on evaluation and assessment in software
  engineering, 2014, pp. 1--10.

\bibitem{garousi2019guidelines}
V.~Garousi, M.~Felderer, M.V. M{\"a}ntyl{\"a}, Guidelines for including grey
  literature and conducting multivocal literature reviews in software
  engineering, Information and Software Technology 106 (2019) 101--121.

\bibitem{braun2006using}
V.~Braun, V.~Clarke, Using thematic analysis in psychology, Qualitative
  research in psychology 3~(2) (2006) 77--101.

\bibitem{sbaraini2011grounded}
A.~Sbaraini, S.M. Carter, R.W. Evans, A.~Blinkhorn, How to do a grounded theory
  study: a worked example of a study of dental practices, BMC medical research
  methodology 11~(1) (2011) 128.

\bibitem{codes_description}
M.~Rosala, \href{https://www.nngroup.com/articles/thematic-analysis/}{How to
  analyze qualitative data from ux research: Thematic analysis} (2019).
\newline\urlprefix\url{https://www.nngroup.com/articles/thematic-analysis/}

\bibitem{patton1990qualitative}
M.Q. Patton, Qualitative evaluation and research methods, SAGE Publications,
  inc, 1990.

\bibitem{zhu2016devops}
L.~Zhu, L.~Bass, G.~Champlin-Scharff, Devops and its practices, IEEE Software
  33~(3) (2016) 32--34.

\bibitem{digital.ai}
Digital.ai, \href{https://digital.ai/periodic-table-of-devops-tools}{Periodic
  table of devops}.
\newline\urlprefix\url{https://digital.ai/periodic-table-of-devops-tools}

\bibitem{wettinger2017collaborative}
J.~Wettinger, U.~Breitenb{\"u}cher, M.~Falkenthal, F.~Leymann, Collaborative
  gathering and continuous delivery of devops solutions through repositories,
  Computer Science-Research and Development 32~(3) (2017) 281--290.

\bibitem{humble2010continuous}
J.~Humble, D.~Farley, Continuous delivery: reliable software releases through
  build, test, and deployment automation, Pearson Education, 2010.

\bibitem{jaatun2012hunting}
M.G. Jaatun, Hunting for aardvarks: Can software security be measured?, in:
  International Conference on Availability, Reliability, and Security,
  Springer, 2012, pp. 85--92.

\bibitem{owasp}
OWASP,
  \href{https://owasp.org/www-community/Application_Threat_Modeling}{Application
  threat modeling}.
\newline\urlprefix\url{https://owasp.org/www-community/Application_Threat_Modeling}

\bibitem{lemos}
R.~Lemos,
  \href{https://techbeacon.com/app-dev-testing/app-sec-service-ready-fast-lane}{App
  sec as a service: Ready for the fast lane?}
\newline\urlprefix\url{https://techbeacon.com/app-dev-testing/app-sec-service-ready-fast-lane}

\bibitem{tahaei2019survey}
M.~Tahaei, K.~Vaniea, A survey on developer-centred security, in: 2019 IEEE
  European Symposium on Security and Privacy Workshops (EuroS\&PW), IEEE, 2019,
  pp. 129--138.

\bibitem{Gitlab2020}
Gitlab, {Mapping the DevSecOps Landscape}, Tech. rep. (2020).

\bibitem{Migues2020}
S.~Migues, J.~Steven, M.~Ware,
  \href{https://www.bsimm.com/download.html}{{Building Security In Maturity
  Model (BSIMM) - Version 11}}, Tech. rep. (2020).
\newline\urlprefix\url{https://www.bsimm.com/download.html}

\bibitem{sanchez2018characterizing}
M.~S{\'a}nchez-Gord{\'o}n, R.~Colomo-Palacios, Characterizing devops culture: a
  systematic literature review, in: International Conference on Software
  Process Improvement and Capability Determination, Springer, 2018, pp. 3--15.

\bibitem{Atlassian.com}
Atlassian.com,
  \href{https://www.atlassian.com/team-playbook/examples/devops-culture}{{Building
  a DevOps culture}}.
\newline\urlprefix\url{https://www.atlassian.com/team-playbook/examples/devops-culture}

\bibitem{unterkalmsteiner2011evaluation}
M.~Unterkalmsteiner, T.~Gorschek, A.K.M.M. Islam, C.K. Cheng, R.B. Permadi,
  R.~Feldt, Evaluation and measurement of software process improvement—a
  systematic literature review, IEEE Transactions on Software Engineering
  38~(2) (2011) 398--424.

\bibitem{khatibsyarbini2018test}
M.~Khatibsyarbini, M.A. Isa, D.N.A. Jawawi, R.~Tumeng, Test case prioritization
  approaches in regression testing: A systematic literature review, Information
  and Software Technology 93 (2018) 74--93.

\bibitem{zhou2016map}
X.~Zhou, Y.~Jin, H.~Zhang, S.~Li, X.~Huang, A map of threats to validity of
  systematic literature reviews in software engineering, in: 2016 23rd
  Asia-Pacific Software Engineering Conference (APSEC), IEEE, 2016, pp.
  153--160.

\end{thebibliography}

\section{Appendix: List of selected papers}
\footnotesize

\begin{itemize}
    
\item [P01] J. Bj{\o}rgeengen, A Multitenant Container Platform with OKD, Harbor Registry and ELK, International Conference on High Performance Computing, 2019
\item [P02] V. Casola, A. De Benedictis, M. Rak and U. Villano, A novel Security-by-Design methodology: Modeling and assessing security by SLAs with a quantitative approach, Journal of Systems and Software, 2020
\item [P03] S.B.O. Gennari Carturan and D.H. Goya, A systems-of-systems security framework for requirements definition in cloud environment, European Conference on Software Architecture (ECSA), 2019
\item [P04] N. Tomas, J. Li, H. Huang, An empirical study on culture, automation, measurement, and sharing of DevSecOps, International Conference on Cyber Security and Protection of Digital Services (Cyber Security), 2019
\item [P05] M.G. Jaatun, Architectural Risk Analysis in Agile Development of Cloud Software, International Conference on Cloud Computing Technology and Science (CloudCom), 2019
\item [P06] V. Mohan, L. ben Othmane and A. Kres, BP: Security concerns and best practices for automation of software deployment processes: An industrial case study, IEEE Cybersecurity Development (SecDev), 2018
\item [P07] E. Zheng, P. Gates-Idem and M. Lavin, Building a virtually air-gapped secure environment in AWS: with principles of DevOps security program and secure software delivery, Annual Symposium and Bootcamp on Hot Topics in the Science of Security, 2018
\item [P08] M.Z. Abrahams and J.J. Langerman, Compliance at Velocity within a DevOps Environment, International Conference on Digital Information Management (ICDIM), 2018 
\item [P09] P. Rimba, L. Zhu, L. Bass,  I. Kuz and S. Reeves, Composing patterns to construct secure systems, European Dependable Computing Conference (EDCC), 2015 
\item [P10] N. Ferry, P.H. Nguyen, H. Song, E. Rios, E. Iturbe, S. Martinez and A. Rego, Continuous Deployment of Trustworthy Smart IoT Systems, Journal of Object Technology 2020
\item [P11] T. M{\aa}rtensson, D. St{\aa}hl and J. Bosch, Continuous Integration applied to software-intensive embedded systems: problems and experiences, International Conference on Product-Focused Software Process Improvement (PROFES), 2016
\item [P12] K. Vijayakumar and C. Arun, Continuous security assessment of cloud based applications using distributed hashing algorithm in SDLC, Cluster Computing, 2019 
\item [P13] H.M. von Stockhausen and M. Rose, Continuous security patch delivery and risk management for medical devices, IEEE International Conference on Software Architecture Companion (ICSA-C), 2020 
\item [P14] B. Fitzgerald and K.J. Stol, Continuous software engineering and beyond: trends and challenges, International Workshop on Rapid Continuous Software Engineering (RCoSE), 2014
\item [P15] J. Brewer, G. Joyce and S. Dutta, Converging Data with Design Within Agile and Continuous Delivery Environments, International Conference of Design, User Experience, and Usability, 2017
\item [P16] X. Larrucea, A. Berreteaga, and I. Santamaria, Dealing with security in a real DevOps environment, European Conference on Software Process Improvement, 2019
\item [P17] M. Lescisin, Q.H. Mahmoud and A. Cioraca, Design and Implementation of SFCI: A Tool for Security Focused Continuous Integration, Computers, 2019
\item [P18] I. Weber, S. Nepal and L. Zhu, Developing dependable and secure cloud applications, IEEE Internet Computing, 2016 
\item [P19] N. Ferry, J. Dominiak, A. Gallon, E. Gonz\'alez, E. Iturbe, S. Lavirotte, S. Martinez, A. Metzger, V. Munt\'es-Mulero, P. H. Nguyen, A. Palm, A. Rego, E. Rios, D. Riviera, A. Solberg, H. Song, J. Y. Tigli and T. Winter, Development and Operation of Trustworthy Smart IoT Systems: The ENACT Framework, International Workshop on Software Engineering Aspects of Continuous Development and New Paradigms of Software Production and Deployment, 2019
\item [P20] M.G. Jaatun, D.S. Cruzes and J. Luna, DevOps for better software security in the cloud invited paper, International Conference on Availability, Reliability and Security, 2017
\item [P21] M. Zaydi and B. Nassereddine, DevSecOps Practices for an Agile and Secure IT Service Management, Journal of Management Information and Decision Sciences, 2019
\item [P22] K. Brady,  S. Moon, T. Nguyen and  J. Coffman, Docker container security in cloud computing, Computing and Communication Workshop and Conference (CCWC), 2020
\item [P23] A. Martin, S. Raponi, T. Combe and R. Di Pietro, Docker ecosystem – Vulnerability Analysis, Computer Communications, 2018
\item [P24] T.Q. Thanh, S. Covaci, T. Magedanz, P. Gouvas and A. Zafeiropoulos, Embedding security and privacy into the development and operation of cloud applications and services, International Telecommunications Network Strategy and Planning Symposium (Networks), 2016
\item [P25] J.A. Kupsch, B.P. Miller, V. Basupalli and J. Burger, From Continuous Integration to Continuous Assurance, IEEE Annual Software Technology Conference (STC), 2017
\item [P26] F. Zampetti, S. Scalabrino, R. Oliveto, G. Canfora, and M. Di Penta, How open source projects use static code analysis tools in Continuous Integration pipelines, International Conference on Mining Software Repositories (MSR), 2017
\item [P27] J.A. Morales, H. Yasar, and A. Volkman, Implementing DevOps practices in highly regulated environments, International Workshop on Secure Software Engineering in DevOps and Agile Development (SecSE), 2018.
\item [P28] T. Soenen, S. Van Rossem, W. Tavernier, F. Vicens, D. Valocchi, P. Trakadas, P. Karkazis, G. Xilouris, P. Eardley, S. Kolometsos, M. Kourtis, D. Guija, S. Siddiqui, P. Hasselmeyer, J. Bonnet and D. Lopez, Insights from SONATA: Implementing and Integrating a Microservice-based NFV Service Platform with a DevOps Methodology, IEEE/IFIP Network Operations and Management Symposium, 2018
\item [P29] M. Ahmadvand, A. Pretschner, K. Ball and D. Eyring, Integrity protection against insiders in microservice-based infrastructures: From threats to a security framework, Federation of International Conferences on Software Technologies: Applications and Foundations (STAF), 2018
\item [P30] A. Khan, Key characteristics of a container orchestration platform to enable a modern application, IEEE Cloud Computing, 2017
\item [P31] K. Brown and C. Hay, Patterns of software development with containers, Conference on Pattern Languages of Programs, 2018 
\item [P32] G. Siewruk, W. Mazurczyk, and A. Karpi\'nski, Security Assurance in DevOps Methodologies and Related Environments, International Journal of Electronics and Telecommunications, 2019
\item [P33] N. Wilde, B. Eddy, K. Patel, N. Cooper, V. Gamboa, B. Mishra and K. Shah, Security for DevOps Deployment Processes: Defenses, Risks Research Directions, International Journal of Software Engineering \& Applications, 2016
\item [P34] V. Gruhn, C. Hannebauer and C. John, Security of public continuous integration services, International Symposium on Open Collaboration (OpenSym), 2013
\item [P35] M. Atighetchi, B. Simidchieva and K. Olejnik, Security Requirements Analysis - A Vision for an Automated Toolchain, International Conference on Software Quality, Reliability and Security Companion (QRS-C), 2019
\item [P36] F. Ullah, A. J. Raft, M. Shahin, M. Zahedi and M.A. Babar, Security support in continuous deployment pipeline, International Conference on Evaluation of Novel Approaches to Software Engineering, 2017
\item [P37] E. Rios, E. Iturbe, and M.C. Palacios, Self-healing multi-cloud application modeling, International Conference on Availability, Reliability and Security, 2017
\item [P38] J. D{\'\i}az,  J.E. P{\'e}rez, M.A. Lopez-Pe{\~n}a, G.A. Mena and A. Yag{\"u}e, Self-Service Cybersecurity Monitoring as Enabler for DevSecOps, IEEE Access, 2019
\item [P39] E. Rios, E. Iturbe, X. Larrucea, M. Rak, W. Mallouli, J. Dominiak, V. Munt\'es, P. Matthews and L. Gonzalez, Service level agreement-based GDPR compliance and security assurance in (multi) cloud-based systems, IET Software, 2019
\item [P40] J.M. Schleicher, M. V{\"o}gler, C. Inzinger and S. Dustdar, Smart brix - a continuous evolution framework for container application deployments, Peer J Computer Science, 2016
\item [P41] M.G. Jaatun, Software security activities that support incident management in secure DevOps, International Conference on Availability, Reliability and Security, 2018
\item [P42] A.A.U. Rahman and L. Williams, Software security in DevOps: synthesizing practitioners' perceptions and practices, International Workshop on Continuous Software Evolution and Delivery (CSED), 2016
\item [P43] A. Rahman, C, Parnin and L. Williams, The seven sins: Security smells in Infrastructure as Code scripts, International Conference on Software Engineering (ICSE), 2019
\item [P44] T. Combe,  A. Martin and R. Di Pietro, To docker or not to docker: A security perspective, IEEE Cloud Computing, 2016
\item [P45] S. Rafi, W. Yu and M.A. Akbar, Towards a hypothetical framework to secure DevOps adoption: Grounded theory approach, Evaluation and Assessment in Software Engineering (EASE), 2020
\item [P46] K. Kritikos, M. Papoutsakis, S. Ioannidis and K. Magoutis, Towards Configurable Cloud Application Security, International Symposium on Cluster, Cloud and Grid Computing (CCGRID), 2019
\item [P47] A. Steffens, H. Lichter and M. Moscher, Towards Data-driven Continuous Compliance Testing, Workshop on Continuous Software Engineering, 2018
\item [P48] M. Guerriero, D.A. Tamburri, Y. Ridene, F. Marconi, M. M. Bersani and M. Artac, Towards DevOps for Privacy-by-Design in Data-Intensive Applications: A Research Roadmap, International Conference on Performance Engineering Companion, 2017
\item [P49] M. Hilton, N. Nelson, T. Tunnell, D. Marinov, D. Dig, Trade-offs in continuous integration: assurance, security, and flexibility, Joint Meeting of the European Software Engineering Conference and the ACM SIGSOFT Symposium on the Foundations of Software Engineering, 2017
\item [P50] C. Paule, T.F. D{\"u}llmann,  A.V. Hoorn, Vulnerabilities in Continuous Delivery Pipelines? A Case Study, International Conference on Software Architecture Companion (ICSA-C), 2019
\item [P51] A. Rahman, A. Partho, P. Morrison, and L. Williams, What questions do programmers ask about configuration as code?, International Workshop on Rapid Continuous Software Engineering, 2018
\item [P52] H. Yasar and K. Kontostathis, Where to integrate security practices on DevOps platform, International Journal of Secure Software Engineering (IJSSE), 2016
\item [P53] S. Faily and C. Iacob, Design as code: Facilitating collaboration between usability and security engineers using cairis, IEEE 25th International Requirements Engineering Conference Workshops (REW), 2017
\item [P54] D. Trihinas, A Tryfonos, M.D. Dikaiakos,  and G. Pallis,  Devops as a service: Pushing the boundaries of microservice adoption, IEEE Internet Computing, 2018

\end{itemize}

\end{document}